\documentclass[aip,pof,preprint,amsmath,amssymb]{revtex4-1}

\usepackage{graphicx}
\usepackage{dcolumn}
\usepackage{bm}
\usepackage[utf8]{inputenc}
\usepackage[T1]{fontenc}
\usepackage{etoolbox}

\usepackage{physics}
\usepackage[ruled,lined]{algorithm2e}
\usepackage{hyperref}

\makeatletter
\def\@email#1#2{
	\endgroup
	\patchcmd{\titleblock@produce}
	{\frontmatter@RRAPformat}
	{\frontmatter@RRAPformat{\produce@RRAP{*#1\href{mailto:#2}{#2}}}\frontmatter@RRAPformat}
	{}{}
}
\makeatother
\begin{document}
	
\preprint{AIP/123-QED}

\title[Modal analysis of blood flows in saccular aneurysms]{Modal analysis of blood flows in saccular aneurysms}

\author{Thien-Tam Nguyen}
\affiliation{
	Department of Civil, Construction, and Environmental Engineering \\
	North Dakota State University \\
	Fargo, North Dakota, United States, 58102.
}
				
\author{Davina Kasperski}
\affiliation{ 
	Department of Civil, Construction, and Environmental Engineering \\
	North Dakota State University \\
	Fargo, North Dakota, United States, 58102.
}

\author{Phat Kim Huynh}
\affiliation{ 
	Department of Industrial and Management Systems Engineering \\
	University of South Florida \\
	Tampa, Florida, United States, 33620.
}

\author{Trung Quoc Le}
\affiliation{
	Department of Industrial and Management Systems Engineering \\
	University of South Florida \\
	Tampa, Florida, United States, 33620.
}

\author{Trung Bao Le}
\email{trung.le@ndsu.edu.}
\homepage{https://sites.google.com/view/complexfluidlab/}
\affiliation{
	Department of Civil, Construction, and Environmental Engineering \\
	North Dakota State University \\
	Fargo, North Dakota, United States, 58102.
}

\date{\today}

%%=================================%
%----------- ABSTRACT -------------%
%==================================%
\begin{abstract}
Currently, it is challenging to investigate aneurismal hemodynamics based on current \textit{in-vivo} data such as Magnetic Resonance Imaging or Computed Tomography due to the limitations in both spatial and temporal resolutions. In this work, we investigate the use of modal analysis at various resolutions to examine its usefulness for analyzing blood flows in brain aneurysms. Two variants of Dynamic Mode Decomposition (DMD): (i) Hankel-DMD; and (ii) Optimized-DMD, are used to extract the time-dependent dynamics of blood flows during one cardiac cycle. First, high-resolution hemodynamic data in patient-specific aneurysms are obtained using Computational Fluid Dynamics. Second, the dynamics modes, along with their spatial amplitudes and temporal magnitudes are calculated using the DMD analysis. Third, an examination of DMD analyses using a range of spatial and temporal resolutions of hemodynamic data to validate the applicability of DMD for low-resolution data, similar to ones in clinical practices. Our results show that DMD is able to characterize the inflow jet dynamics by separating large-scale structures and flow instabilities even at low spatial and temporal resolutions. Its robustness in quantifying the flow dynamics using the energy spectrum is demonstrated across different resolutions in all aneurysms in our study population. Our work indicates that DMD can be used for analyzing blood flow patterns of brain aneurysms and is a promising tool to be explored in \textit{in-vivo}.
\end{abstract}

\maketitle

%%=================================%
%--------- INTRODUCTION -----------%
%==================================%
\section{Introduction}
\label{sec:introduction}
Enlargements of blood vessel walls in the brain known as intracranial aneurysms (IAs) run a great risk of rupturing, which would cause major medical problems \cite{gholampour2021effect, rousseau2021location}. Severe results from a rupture of IA could include hemorrhage, stroke, brain damage, or death \cite{toth2018, lauzier2023early, hoh20232023}. In addition, these aneurysms affect between 2\% and 5\% of individuals worldwide \cite{thompson2015}. Remarkably, many aneurysms show no symptoms before rupture and are often discovered by chance during medical imaging for other unrelated illnesses \cite{khanetal2021, mayoBrainSymptoms2022, tawk2021diagnosis}. Clinical imaging techniques have found that most IAs occur along the Circle of Willis (CoW) \cite{feng2023anatomical}, a vital structure that distributes blood to various parts of the brain. Once identified, therapeutic options, including surgical clipping, endovascular coiling, and flow diverter, \cite{rayzco2020} are evaluated based on a meticulous risk evaluation of the likelihood of rupture compared to the hazards associated with surgery. The complexity of these assessments emphasizes the critical need for consistent risk predictors to help physicians effectively control unruptured IAs. 

Existing evaluation methods that combine aneurysm parameters, i.e., location, size, and form, with patient history only provide a crude estimate of stability \cite{alwalid2022artificial, tawk2021diagnosis, tang2022morphological}. Despite clinical efforts, the risk prediction methods used in practice for brain aneurysms are not fully established, mainly because there is still an insufficient understanding of the pathophysiology that initiates arterial enlargement and aneurysm progression. Even the most effective scoring techniques, such as the ELAPSS (Earlier Subarachnoid Hemorrhage, Location of the Aneurysm, Age $>60$ Years, Population, Size of the Aaneurysm, and Shape of the Aneurysm) and the PHASES (Population, Hypertension, Age, Size, Earlier Subarachnoid Hemorrhage), exhibit inadequate predictive accuracy (less than 20\%) \cite{hilditchAppPHASES2021, khanetal2021, rayzco2020}. Although there have been improvements in neuroimaging and surgical procedures, accurately assessing the likelihood of aneurysm rupture continues to be a difficult task. This challenge emphasizes the need to identify patient-specific key markers to improve predictive models for aneurysm growth and rupture. More sophisticated approaches incorporate dynamic factors like blood flow patterns and Wall Shear Stress (WSS), which can be observed by high-resolution imaging and computer modeling \cite{rayzco2020, zingaro2024comprehensive, sheikh2020review}. Overall, these methods are designed to improve the accuracy of predictions, enabling more customized and efficient treatment procedures.

The emergence of incorporating the WSS into current imaging techniques reflects the elevated prevalence of the IAs in the CoW \cite{BROWN2014393}. This implies that hemodynamic factors are fundamental in their pathophysiology and rupture, which has been hypothesized in the literature by both empirical and computational studies \cite{sforza2009hemodynamics, rayzco2020, schnell2014three, schnell20164d, dholakiaetAl2017}. Blood flows in brain aneurysms exhibit chaotic and transient states, often with vortices \cite{sforza2009hemodynamics, xiangetAl2013, wuetal2021hemo}, unlike the laminar flow in healthy brain arteries. These "disturbed" flow patterns have fast-moving inflow jets and vortex ring structures \cite{le2010pulsatile} that can collide with the aneurysm's wall, resulting in abnormally high WSS \cite{sforza2009hemodynamics, wuetal2021hemo}. It prompted clinicians to look for hemodynamic measures \citep{futami2019identification}, including the time-averaged wall shear stress (TAWSS) and the oscillatory shear index (OSI), as potential indicators of rupture risk \cite{wuetal2021hemo}. 

Nonetheless, evaluating the WSS values and its gradient presents challenges with \textit{in-vivo} data \cite{wuetal2021hemo, khanetal2021, macdonaldetal2022}. Contemporary state-of-the-art MRI and CT measures possess temporal resolutions of approximately 40 milliseconds and spatial resolutions exceeding 1 millimeter \cite{schnell2014three}. Still, these resolutions are too low for \textit{in-vivo} WSS estimates \cite{wuetal2021hemo}. While \textit{in-vivo} technologies slowly improve and approach sufficient precision to provide accurate WSS, the transient and chaotic nature of aneurysmal flows \cite{le2010pulsatile} complicate the interpretation of \textit{in-vivo} data. Consequently, there is an immediate necessity to develop methods that can interpret low-resolution \textit{in-vivo} datasets \cite{habibi2021integrating}. Addressing this gap is crucial for advancing the diagnostic and prognostic capabilities of neuroimaging in the context of IAs.

The amalgamation of machine learning (ML) and sensor-based, data-driven approaches with fluid dynamics has significantly transformed the analysis of cardiovascular flows. It introduces fresh interpretations of \textit{in-vivo} data \cite{Arzani.2020, arzani2022machine}. Notable among these advancements are data reduction techniques such as Proper Orthogonal Decomposition (POD) and Dynamic Mode Decomposition (DMD), which have become influential tools for collecting and evaluating dynamic flow patterns \cite{taira2017modal}. These unsupervised and equation-free techniques are highly effective in finding important flow features, such as flow instabilities, vortex interactions, and other complex fluid phenomena \cite{taira2017modal, arzani2022machine}. Both POD and DMD methods analyze the dynamics of blood flow by breaking it down into separate modal structures, known as "spatial modes". However, DMD can identify the modal regions with specific frequencies of flow fluctuation. The range of these modes extends from low frequencies, which are linked to laminar flow (less than or equal to 15 Hz), to higher frequencies indicating flow instabilities \cite{LE2021110238, yu2022application}.

The application of DMD in both \textit{in-vitro} and \textit{in-silico} flow measurements has demonstrated its proficiency in managing noisy data and its capability to uncover complex dynamic patterns in aneurysmal flow. A study conducted by Le (2021) \cite{LE2021110238} shows that the application of the Exact DMD method to two patient-specific simulation data yields normalized Cumulative Energy (CE) curves that differentiate the datasets into two distinct patterns, signifying varying flow dynamics. In another recent study employing POD on various flow scenarios, Csala et al. \cite{csala2022comparing} note the CE curve from the case exhibiting laminar flow requires less mode to accumulate to 100\% of total energy. When comparing two cases of aneurysmal flow at Reynolds numbers $Re = 432$ and $Re = 322$, the latter case shows a slower decay in the singular values despite having a smaller Reynolds number. The findings suggest that CE curves, derived from POD and DMD methodologies, could serve as a novel metric for evaluating aneurysmal flows. Ultimately, while previous studies \cite{LE2021110238, csala2022comparing, yu2022application} focused on evaluating DMD's capacity to quantify flow dynamics, there has been an absence of attempts to show DMD's dependability and viability for a heterogeneous cohort of the patient population.

The main goal of this work is to verify the efficacy of DMD in stratifying blood flow patterns from patient-specific aneurysms. Specifically, we study two variants: (1) the Hankel DMD \cite{doi:10.2514/1.J056060, kamb2020, pan2020}, which incorporates time-delay embedding \cite{citeulike:2735031}; and (2) the Optimized DMD \cite{askham2018variable}, which involves minimizing the residual of the DMD projection. After collecting data from the cohort population, we aim to infer blood flow dynamics in IAs by employing the following patient-specific framework comprised of the following 5 stages:
\begin{enumerate}
	\item \textit{High-resolution simulations:} Perform Computational Fluid Dynamic (CFD) simulations to obtain detailed flow data that reflect the unique characteristics of each patient's aneurysm;
	\item \textit{Modal analysis:} Apply Hankel DMD and Optimized DMD to analyze the high-resolution flow data, highlighting its capacity to discern key flow features;
	\item \textit{Patient stratification:} Evaluate the capability of the Hankel DMD method in detecting critical flow instabilities, such as the transition to turbulence, which are vital in the aid of predicting rupture risks;
	\item \textit{Robustness assessment:} Conduct a detailed examination of DMD's sensitivity to its resolution dependency, critically assessing its reliability in clinical settings; and
	\item \textit{Comparative analysis:} Compare results from the Hankel DMD and the Optimized DMD methods from the same patient-specific data to quantify the difference brought to the DMD kernel by each modification.
\end{enumerate}

Our paper's forward is organized into several sections. In Sec. \ref{sec:methods}, the criteria for selecting patient cases are presented, accompanied by morphological descriptions of each aneurysm. In addition, this section also delineates the 3D modeling process and simulation details. In sections \ref{sec:results} and \ref{sec:discussion}, the CFD and DMD results are presented and discussed in the context of our patient-specific framework. Finally, conclusions are encapsulated in Sec. \ref{sec:conclusions}.

%%=================================%
%----------- METHODS --------------%
%==================================%
\section{Materials and Methods}
\label{sec:methods}
\subsection{Patient-specific Aneurysmal Anatomies}
\label{subsec:aneu_models}
\subsubsection{3D Anatomical Models}
\begin{figure}
	\centering
	\includegraphics[width=\textwidth]{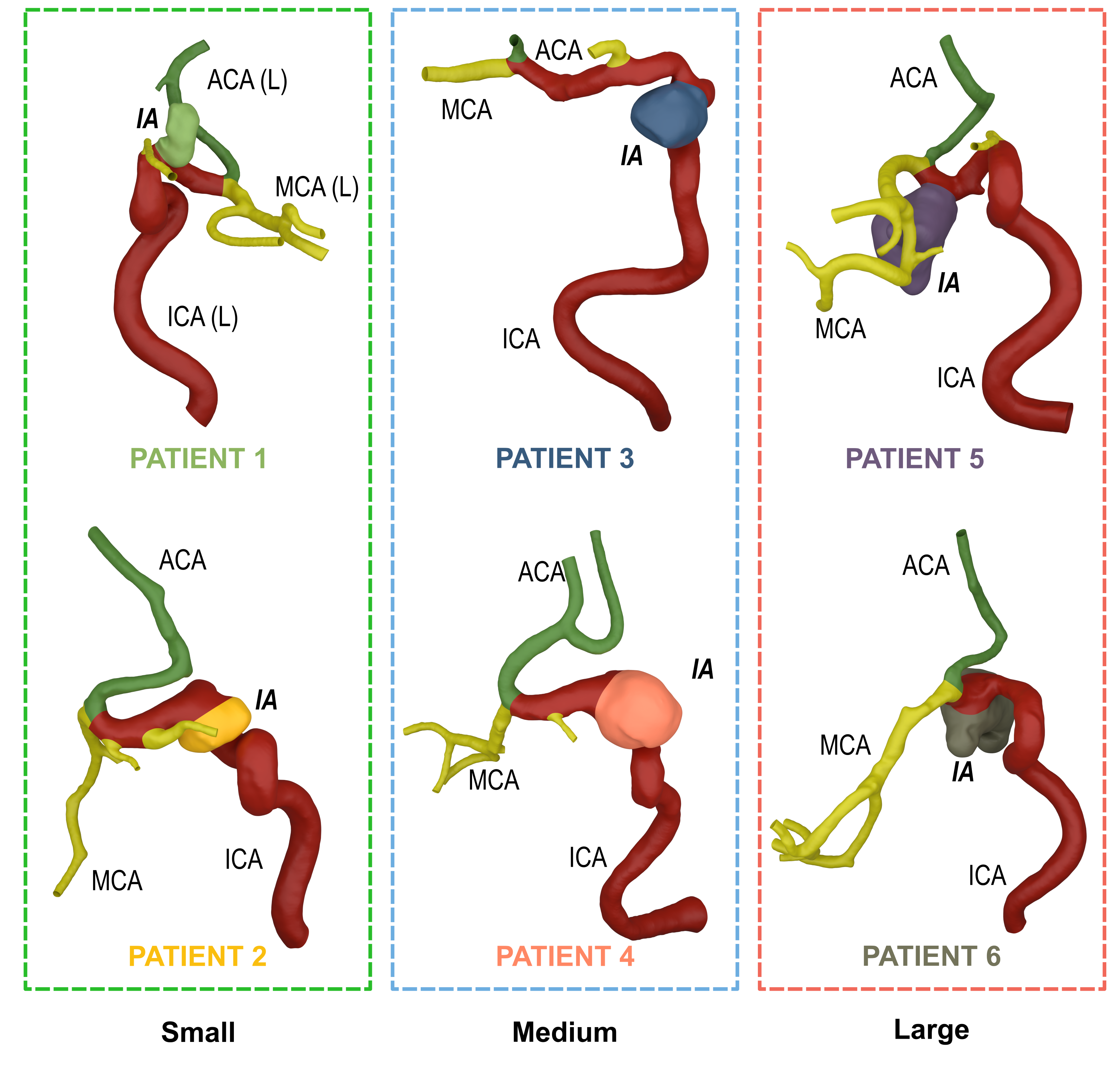}
	\caption{Anatomical geometries of six patient-specific aneurysms reconstructed from DICOM images in the \textit{Aneurisk project}. The geometric details of each aneurysm are shown in Table \ref{tab:anatomical-parameters}. The locations of the Internal Carotid Artery (ICA), Middle Cerebral Artery (MCA), and Anterior Communicating Artery (ACA) segments are color-coded. The general flow direction is from the caudal to the cranial (heart to brain direction). The aneurysm sacs for patients 1, 2, 3, 4, 5, and 6 are color-coded (Green, Yellow, Orange, Purple, Dark Blue, and Brown) throughout the presentations in this work. The data related to the patients in subsequence figures use similar color codes.}
	\label{fig:geom}
\end{figure}

IAs are typically categorized into two basic shapes: (1) saccular, or (2) fusiform \cite{rayzco2020}. We focus on saccular aneurysms, which are more commonly found in the brain \cite{frosen2012}. In this work, the patient-specific imaging data of brain aneurysms was selected from the repository of Digital Imaging and Communications in Medicine (DICOM) files from the \href{https://github.com/permfl/AneuriskData}{AneuRisk project} \cite{aneurisk2011}. Using a retrospective approach, the study cohort was formed from six aneurysms situated in the Internal Carotid Artery (ICA). An ICA aneurysm is the most common IA and its formation is linked to the bifurcation in the interconnected segment with the Middle Cerebral Artery (MCA) and the Anterior Communicating Artery (ACA). Based on the Bouthillier categorization \cite{icasegment}, all six aneurysms occur in the vessel wall between the ophthalmic (supraclinoid) (C6) and communicative (terminal) (C7) segments of their ICAs (see Fig. \ref{fig:geom}). Specifically, the aneurysms of patients 1, 2, 3, and 4 were developed on the endothelial wall of the ICA artery at C6, while patients 5 and 6 had their aneurysms at C7. 

\href{https://www.slicer.org/}{3D Slicer} \cite{FEDOROV20121323}, an open-source software, was used for processing the DICOM images and reconstructing the 3D anatomical models. Segmentation is performed first and it yields the CoW and various brain artery segments, which are exported in stereolithography (STL) file format. Next, the model goes through several post-processing steps in another open-source software, \href{http://www.meshmixer.com}{Meshmixer} (RRID:SCR\_015736). In those steps, the CoW is isolated from the attached arteries and then cleaned up. A technique is carried out to smooth the arterial wall by re-meshing the model surfaces. Fig. \ref{fig:geom} shows the final 3D models of all 6 patients from the segmentation and post-processing steps that only include the surfaces of the aneurysm and the ICA, the ACA, and the MCA. Finally, each surface mesh was converted into a solid mesh having around 50,000-100,000 vertices. The resulting models after reconstruction from the DICOM images are referred to as anatomical models.

For each aneurysm, three morphological attributes, i.e., artery diameter $D$, aneurysm diameter $W$, and aneurysm height $H$, were measured. Table \ref{tab:anatomical-parameters} summarizes the anatomical attributes of the IAs. The morphological trio was used to calculate several aneurysmal indices, i.e, the Aspect Ratio $AR$:
\begin{equation}
    AR = \frac{W}{H}
\end{equation}
and the Aneurysm Number $An$:
\begin{equation}
    An = PI \frac{W}{D}; 
\end{equation}
where the Pulsatility Index $PI$ is calculated from the velocity profile at the inlet such that:
\begin{equation}
    PI = \frac{U_{\text{max}} - U_{\text{min}}}{\bar{U}}
\end{equation}
where $U$ is the bulk velocity prescribed at the inlet.

The location of each aneurysm ostium was identified by following the 4-point technique from the Anatomy-Guided Multi-Level Exploration \cite{ostiumAnatomyGuide}. Then, the central aneurysm axis (CAS) can be defined as a line perpendicular to the ostium at its center point. The angle of inclination \cite{aneurysmAngle}, or aneurysm angle, is measured as the angle between the incoming flow and the CAS. In addition, the aneurysms can be further classified as terminal or lateral \cite{sforza2009hemodynamics} by measuring an angle between the directions of incoming and exiting flows.

\subsubsection{Hierarchical Clustering}
A classification was carried out using anatomical features of the aneurysms and their parent's ICAs. This is a prerequisite step to investigate the correlation between aneurysm size and its hemodynamics. The Unweighted Pair-Group Method with Arithmetic mean (UPGMA)\cite{Sokal1958ASM, TajimaOxford1990} is a generic method that was employed to analyze the morphological attributes in Table \ref{tab:anatomical-parameters}. 

Assuming the attributes are of an orthogonal set that forms a three-dimensional space, then the location of each individual is a vector $\left[ D_j, W_j, H_j \right]^\intercal$. A matrix of all positions has the following form:
\begin{equation}
    \mathbf{S} = \left[\mathbf{s}_1 \quad \mathsf{s}_2 \: \cdots \: \mathsf{s}_k \: \cdots \: \mathsf{s}_P \right] = \begin{bmatrix}
        D_1 & D_2 & \cdots & D_k & \cdots & D_P \\
        W_1 & W_2 & \cdots & W_k & \cdots & W_P \\
        H_1 & H_2 & \cdots & H_k & \cdots & H_P
    \end{bmatrix} 
\end{equation} 
with $k \in \mathbb{N} \:| \: k = [1, P]$ where $P$ is the total number of individuals. $P = 6$ in our study, which makes the matrix $\mathbf{S}$ a 3-by-6 matrix.

Pairwise Euclidean distance between an arbitrary pair $(i,j)$ of the corresponding column vectors in the matrix $\mathbf{S}$ is calculated as:
\begin{equation}
    d^2_{ij} = (\mathsf{s}_i - \mathsf{s}_j)(\mathsf{s}_i - \mathsf{s}_j)^\intercal
\end{equation}

The UPGMA method creates a simple agglomerative (bottom-up) hierarchical clustering by determining how objects in the dataset should be grouped into clusters. The distance between two clusters $r$ and $t$ is referred as a linkage:
\begin{equation}
    d(r,t) = \frac{1}{n_r n_t} \sum^{n_r}_{i = 1} \sum^{n_t}_{j = 1} d^2_{ij}
\end{equation}
where $n$ is the number of individuals in a cluster.

\subsection{Computational Methodologies}
\label{subsec:comp_details}
\begin{figure}
	\centering
	\includegraphics[width=0.9\textwidth]{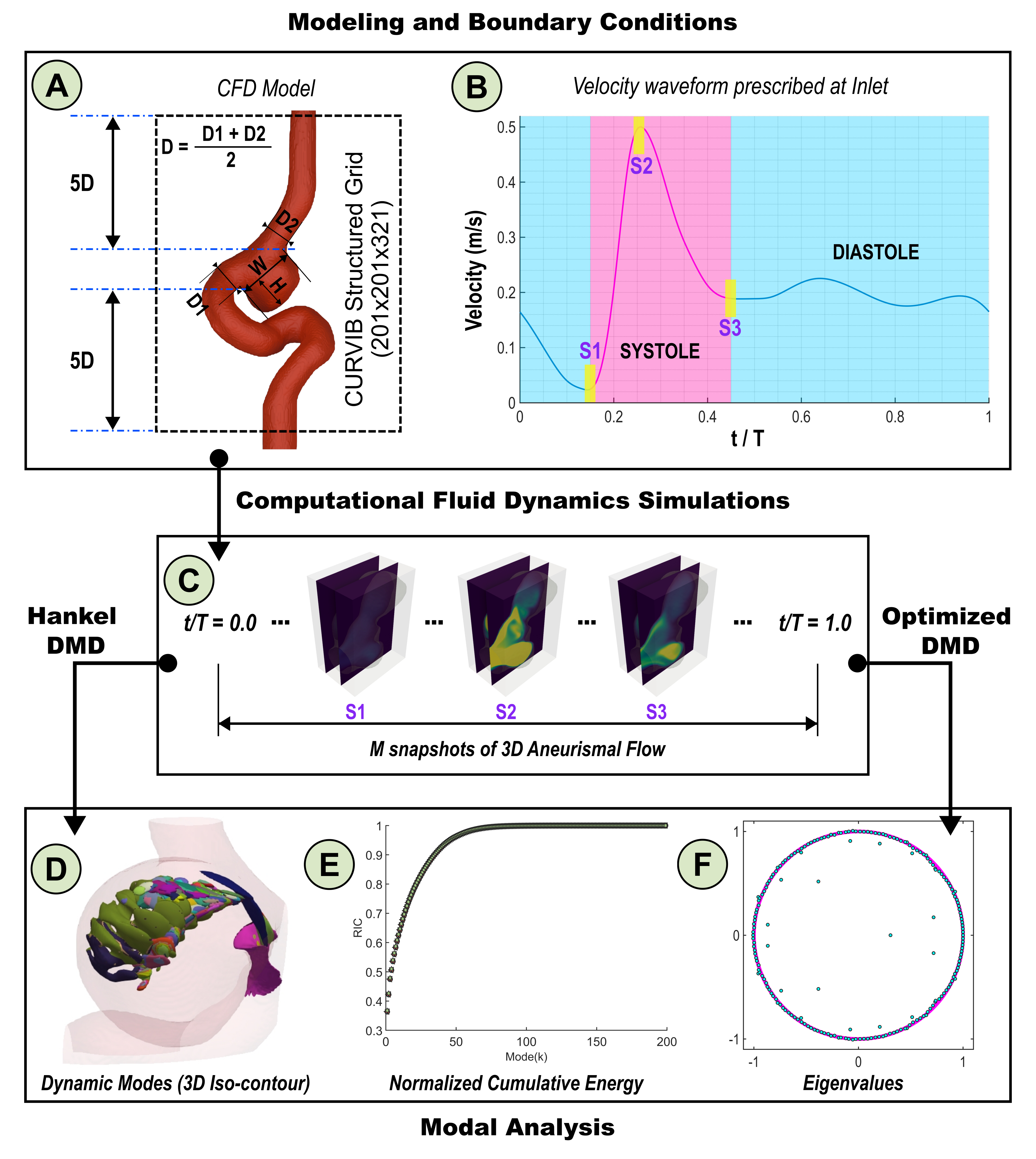}
	\caption{Patient-specific modeling pipeline \citep{review-rayz2020} consists of (1) Image segmentation; (2) Modeling assumptions and flow boundary conditions; (3) Blood flow simulation; and (4) Post-processing for computing clinically relevant metrics. Details in this work are (A) Anatomical geometry and the computational domain; (B): Prescribed flow waveform as at the inlet (ICA). The time instances $S1, S2$, and $S3$ denote the beginning of systole, peak systole, and the end of systole, respectively. (C) Snapshots of CFD simulations are stacked as $N$ instances for Hankel and Optimized DMD analyses; The DMD results include (D) the 3D spatial extents of DMD modes; (E) the cumulative energy spectrum; and (F) the eigenvalues of DMD modes.}
	\label{fig:setup}
\end{figure}

The computational (CFD) models of the aneurysms were created by removing the ACA and the MCA branches. The process was carried out by employing the Plane Cut feature of the Meshmixer software at the end of the C7 segment. The cut surface was elongated in the streamwise direction to a distance of $5D$. The end of the extrusion part serves as the outlet in our simulations. Figure \ref{fig:setup}A illustrates an example of the CFD model of Patient 4's aneurysm after conversion from the anatomical model as a solid body with closed surfaces.

An important characteristic of the anatomical models (refer to Fig. \ref{fig:geom}) is that the segments from the petrous (C2) to the clinoid (C5) exhibit patient specificity. Piccinelli et al. \cite{aneuriskMorphology} show that the ICA geometry, characterized by torsions and bends, strongly influences aneurysmal hemodynamics. Therefore, the conversion step kept the ICA intact. The inlet is designated in the initial section of the ICA. 

A structured grid was created in the commercial software Pointwise \cite{Steinbrenner1991TheG3} as the background mesh as required by the curvilinear immersed boundary (CURVIB) method \cite{gilmanov2015numerical,le2010pulsatile}. The dimensions of the structured grid were $201  \times 201 \times 321$ ($\approx 12$ millions of grid points) in the $x$, $y$, and $z$ axes, respectively. A non-uniform distribution among the grid points was used to achieve a denser concentration of grid points in the aneurysm sac. This technique achieved a spatial resolution between $0.07$ mm and $0.15$ mm in the sac volume. Detailed resolutions in each direction for all computational models are denoted as $1\mathsf{X}$ in Table \ref{tab:spatial-resolution}.

Blood in our simulations was modeled as an incompressible Newtonian fluid \cite{bessonovetAl2015} because the diameters of the ICA's arteries are sufficiently large ($3-4.5$ mm) (see Table \ref{tab:anatomical-parameters}). The governing equations for flow dynamics are the three-dimensional incompressible Navier-Stokes equations with the density and the dynamic viscosity of $\rho = 1000~\text{kg}/\text{m}^3$ and $\nu = 3.35 \times 10^{-6}~\text{m}^2/\text{s}$, respectively. The CURVIB method is used to simulate the blood flow dynamics where the continuity and momentum equations of the Navier-Stokes are discretized in a hybrid staggered/non-staggered grid. The discrete equations are integrated in time using a fractional step method\cite{KIM1985}. In detail, a Newton-Krylov solver is used for the momentum equations, and a GMRES solver with a multigrid pre-conditioner is employed for the Poisson equation \cite{ge2007}. Our in-house code has been validated with \textit{in-vitro} experiment \cite{le2013vortex} and \textit{in-vivo} measurement \cite{le2019high}. The computational code has also been applied to various cardiovascular flow problems \cite{LE201341} including intracranial aneurysms \cite{LE2021110238}.

At the inlet, a uniform velocity profile was prescribed using a flow waveform. The waveform represents a cardiac cycle of $72~\text{bpm}$ or the period of $T = 0.83$ seconds\cite{cebraletAl2005}. As shown in Figure \ref{fig:setup}B, the waveform consists of two distinguishable phases: (i) the systole phase; and (ii) the diastole phase. The bulk velocity at peak systole is $U_0 = 50~\text{cm}/\text{s}$. Using the Fourier transform technique, the inflow waveform was found to have three dominant frequencies. The frequencies are listed in increasing order of amplitude as follows: $1.2$ Hz (first peak); $2.4$ Hz (second peak); and $3.6$ Hz (third peak). For the outlet, the Neumann-type boundary condition was prescribed. No-slip and no-flux conditions were prescribed for other artery (endothelial) walls. The cardiac cycle was discretized into 4,000 equispaced steps with a physical timestep of $\Delta t \approx 0.21$ milliseconds. For each patient-specific case, 4 cycles were performed, and the velocity vectors during the $4^{th}$ cycle were extracted to be used in visualization and modal analysis steps (illustrated in Fig. \ref{fig:setup}C).

Two governing parameters for pulsatile blood flow are the peak Reynolds ($Re$) number and the Womersley ($\alpha$) number. They are defined using the peak systole velocity $U_0$ and the inlet artery's diameter $D_0$ as follows: 
\begin{eqnarray}
	Re = \frac{U_0 D_0}{\nu}\\
	\alpha = \frac{D_0}{2} \sqrt{\frac{2 \pi}{T \nu}}    
\end{eqnarray}
The Reynolds number and Womersley number are all approximately consistent for all cases, which are $Re \approx 550$ and $\alpha \approx 3$ respectively, as shown in Table \ref{tab:anatomical-parameters}.

\subsection{Modal analysis of blood flows using Dynamic Mode Decomposition}
\label{subsec:dmd_methods}
Dynamic Mode Decomposition is a data-driven technique that can capture the underlying dynamics of fluid flow from a set of time-resolved data \cite{taira2017modal}. DMD combines several aspects of POD and discrete Fourier transform \cite{chenVariantsDMD2012, rowleySpectral2009, taira2017modal}. DMD splits a high-dimensional spatiotemporal signal into a triplet of spatial structures/modes, eigenvalues, and scalar amplitudes\cite{schmid2022}. In addition, the foundation of DMD has been improved in several works \cite{nedzhibov2022}. In this study, two variants of DMD were used: (i) the Hankel DMD method \cite{kamb2020}, and (ii) the Optimized DMD method \cite{askham2018variable} (refer to workflow in Fig. \ref{fig:setup}D-F).

DMD methods perform on a data matrix $\mathbf{O}$ of observable such that: 
\begin{equation}
	\mathbf{O} = [\mathbf{V_0}, \mathbf{V_1}, ... , \mathbf{V_M}] = \begin{bmatrix}
		| & | &     & | \\
		\vb{u}_N^0 & \vb{u}_N^1 & \cdots & \vb{u}_N^M \\
		| & | &     & | \\
		\vb{v}_N^0 & \vb{v}_N^1 & \cdots & \vb{v}_N^M \\
		| & | &     & | \\
		\vb{w}_N^0 & \vb{w}_N^1 & \cdots & \vb{w}_N^M \\
		| & | &     & | \\
	\end{bmatrix} 
\end{equation}
where $\mathbf{V}_m (\mathbf{u}, \mathbf{v}, \mathbf{w})$ is the flow field at a time instance (snapshot) with $m \in \mathbb{N} \:| \: m = [0, M]$. Here $\mathbf{u}$, $\mathbf{v}$, and $\mathbf{w}$ are the components of the velocity in $x$, $y$, and $z$ direction, respectively. $N$ is the total number of grid points within a Volume of Interest (VOI). And $M+1$ is the number of snapshots.

\subsubsection{Scope of Analysis}
\label{sec:scope}
The dependability and viability of the DMD methods were investigated with consideration of the following parameters:
\begin{enumerate}
	\item The VOIs are subsets of the computational grids. They are set to contain the patient's aneurysm sac. ParaView, an open-source software, was used to create the VOIs in all cases. The flow data in each case was extracted in a text file (CSV) format. The sensitivity of the DMD analysis to the choice of VOIs will be examined in detail in a later section.
	\item In order to assess the influence of spatial resolution on the DMD analysis, a spatial resampling technique was applied when extracting the VOIs with coarsening factors of two ($2\mathsf{X}$), four ($4\mathsf{X}$), and eight ($8\mathsf{X}$) times in all directions, as indicated in Table \ref{tab:spatial-resolution}. At the lowest spatial resolutions (coarsening factor of $8\mathsf{X}$), the input data has a comparable spatial resolution to the ones from the 4D-flow MRI techniques ($\approx 1$ mm).
	\item In order to assess the influence of temporal resolution on the DMD analysis, a temporal resampling technique was performed by skipping the number of input instances. Three scenarios are considered, i.e., $M = 50, ~100, ~\text{and}~200$, (or equivalently at the temporal resolution of  $\Delta \tau = 16.8, ~8.4, ~\text{and}~4.2$ milliseconds).
\end{enumerate}

In summary, the 6 anatomical geometries, together with the spatial coarsening factors of $1\mathsf{X}$ (CFD resolution), $2\mathsf{X}$, $4\mathsf{X}$, and $8\mathsf{X}$ and the temporal resolutions of $M = 50, ~100, ~\text{and}~200$ resulted in a total of 72 standard datasets for modal analysis.

\subsubsection{The Hankel Modification (Hankel DMD)}
\label{subsubsec:hankel_dmd}
In the Exact DMD method \cite{exactDMD2014, askham2018variable}, the data matrix $\mathbf{O}$ is broken into two sequences:
\begin{align}
	\mathbf{X} &= [\mathbf{V_0}, \mathbf{V_1}, ... , \mathbf{V_{M-1}}] \\
	\mathbf{Y} &= [\mathbf{V_1}, \mathbf{V_2}, ... , \mathbf{V_{M}}]
\end{align}

The DMD analysis assumes that the flow dynamics is based on the following formulation:
\begin{equation}
	\mathbf{Y} = \mathbf{A} \mathbf{X}
\end{equation}

The Exact DMD algorithm aims to calculate the "dynamics" matrix $\mathbf{A}$ using the pseudoinverse ($\dagger$) of $\mathbf{X}$:
\begin{equation}
	\mathbf{A} = \mathbf{Y} \mathbf{X}^{\dagger}
	\label{eq:linear-assumption-A}
\end{equation}

The Singular Value Decomposition (SVD) of the matrix $\mathbf{X}$ is carried out as a prerequisite to the calculation of the pseudoinverse matrix $\mathbf{X}^{\dagger}$. The task of finding the matrix $\mathbf{A}$ involves solving an optimization problem using the Frobenius norm as follows:
\begin{equation}
	\text{minimize} \| \mathbf{Y} - \mathbf{L} \mathbf{A} \mathbf{\Sigma} \mathbf{R}^* \|^2_F
	\label{eq:best-fit-A}
\end{equation}
where $\mathbf{L}$ and $\mathbf{R}$ are left- and right-unitary matrices of SVD. The diagonal matrix $\mathbf{\Sigma}$ contains the singular values $\sigma$. Alg. \ref{alg:exactDMD} provides a step-by-step procedure of the Exact DMD method for computing the matrix $\mathbf{A}$ using non-truncated SVD on the matrix $\mathbf{X}$. 

\begin{algorithm}[b]
	\caption{Exact DMD}
    \label{alg:exactDMD}
		\textbf{1.} Approximate $\mathbf{Y} \approx \mathbf{A} \mathbf{X}$.
		
        \textbf{2.} Perform the Singular Value Decomposition (SVD): $\mathbf{X} = \mathbf{L} \mathbf{\Sigma} \mathbf{R}^*$, Using the singular values ($\sigma_i$) to plot the cummulative energy curve
		
        \textbf{3.} Reconstruct the full-rank matrix $\mathbf{A} = \mathbf{L}^* \mathbf{Y} \mathbf{R} \mathbf{\Sigma}^{-1}$ (no truncation)
		
        \textbf{4.} Find the eigenvalues ($\lambda$) and eigenvectors ($\mathbf{K}$) of the dynamic matrix as $\mathbf{A} \mathbf{K} = \mathbf{K} \mathbf{\Lambda}$. 
        
        \textbf{5.} Compute modes $\mathbf{\varphi} = \frac{1}{\| \mathbf{Y} \mathbf{R} \mathbf{\Sigma}^{-1} \mathbf{K} \|_2} \mathbf{Y} \mathbf{R} \mathbf{\Sigma}^{-1} \mathbf{K}$ and the Jovanovi\'{c} amplitudes \cite{jovanovic2014sparsity} $\mathbf{b}$ .
 \end{algorithm} 

The Hankel variant \cite{doi:10.2514/1.J056060, kamb2020, pan2020} involves applying the Exact DMD algorithm to a Hankelized matrix constructed from the original data, rather than directly to the data snapshots shown in the Alg. \ref{alg:exactDMD}. This modification aims to address the pitfall of forward-time bias.

A time-delay embedding dimension $d$ can be defined when constructing the Hankelized matrix $\mathbf{X}_H$, given data matrix $\mathbf{O}$, as:
\begin{equation}
	\mathbf{X}_{H_d} = \begin{bmatrix}
		\mathbf{V_0} & \mathbf{V_1} & \cdots & \mathbf{V_{M-d}} \\
		\mathbf{V_1} & \mathbf{V_2} & \cdots & \mathbf{V_{1+M-d}} \\
		\vdots & \vdots &   & \vdots \\
		\mathbf{V_d} & \mathbf{V_{d+1}} & \cdots & \mathbf{V_{M}} \\
	\end{bmatrix} 
\end{equation}
where $d = 1$ was chosen in our modal analyses. The resulting Hankelized input matrix has the following form:
\begin{equation}
	\mathbf{X}_{H_1} = \begin{bmatrix}
		\mathbf{V_0} & \mathbf{V_1} & \cdots & \mathbf{V_{M-1}} \\
		\mathbf{V_1} & \mathbf{V_2} & \cdots & \mathbf{V_{M}} \\
	\end{bmatrix} 
\end{equation}

To visualize a DMD mode, a three-dimensional structure (iso-contour) is constructed at a threshold $\frac{|\mathbf{V}|}{| \mathbf{V}_{\text{max}} |} = 0.5$ in the ParaView software. In addition to the 3D structures, the corresponding frequencies of these modes are calculated as $\omega_k = \frac{\log(\lambda_k)}{2 \pi}$ for an arbitrary $k^{th}$ DMD mode. Then, the frequency spectrum is scaled by temporal magnitude $| b_k \lambda^M_k |$, instead of the scalar magnitude $| b_k |$ \cite{LE2021110238}. Ultimately, the temporal magnitudes are normalized by the maximum value.
%--------------------------%
%---- OptDMD --------------%
%%------------------------%%
\subsubsection{Optimized DMD}
\label{subsubsec:optimized_dmd}
Optimized DMD \cite{askham2018variable} is introduced to alleviate the limitations of the Exact DMD method when dealing with non-linear phenomena. The method performs the mode decomposition process on the entire data $\mathbf{O}$ with the objective is minimizing the difference between the actual data $\mathbf{O}$ and the data reconstructed from the Optimized DMD modes.

The algorithm starts by re-writing the matrix $\mathbf{O}$ as follows:
\begin{equation}
    \mathbf{O}^\intercal \approx \bm{\Phi}(\bm{\lambda}) \mathbf{B}
\end{equation}

The optimization starts with initial guesses $\bm{\gamma}$ for the eigenvalues $\bm{\lambda}$ and $\bm{\beta}$ for the matrix $\mathbf{B}$. The algorithm continues to refine iteratively by solving an exponential fitting problem such that:
\begin{equation}
	\text{minimize} \, \| \mathbf{O}^\intercal - \mathbf{\Phi}(\bm{\gamma}) \bm{\beta} \|_F \quad \quad \quad \text{over} \quad \mathbf{\gamma} \in \mathbb{C}^k , \bm{\beta} \in \mathbb{C}^{M \times N}
	\label{eq: optdmd_step2}
\end{equation}

After the convergence, the Optimized DMD modes are recovered as: 
\begin{equation}
	\varphi_i = \frac{1}{\| \hat{\mathbf{B}}^\intercal (:,i) \|_2} \hat{\mathbf{B}}^\intercal (:,i)
	\label{eq: optdmd_step3}
\end{equation}
where $\mathbf{B}^\intercal(:,i)$ is the $i^{th}$ column of $\mathbf{B}^\intercal$. And the Optimized DMD amplitudes are:
\begin{equation}
    b_i = \| \hat{\mathbf{B}}^\intercal (:,i) \|_2
\end{equation}

As the optimization is carried out over the entire dataset $\mathbf{O}$, this DMD algorithm significantly enhances the robustness and accuracy of mode decomposition and reduces the impact of noises which is the second disadvantage of the Exact DMD method. In addition, the Optimized DMD algorithm allows us to capture the inherent non-linearity and transient phenomena in flow dynamics. These attributes are essential in accurately assessing flows in IAs.  

%\begin{algorithm}[b]
%	\caption{Optimized DMD}\label{alg:OptimizedDMD}
%	\begin{algorithmic}[1]
%		\State Let the data matrix $\mathbf{O}$ and an initial guess for $\mathbf{\gamma}$ be given.
%		\State Solve the problem defined at Equation \ref{eq: optdmd_step2} using variable projection algorithm.
%		\State Set $\lambda_i = \hat{\gamma}_i$.
%		\State Calculate the eigenmodes defined at Equation \ref{eq: optdmd_step3}.
%		\State Set the amplitude as $b_i = \| \hat{\mathbf{B}}^\intercal (:,i) \|_2$.
%	\end{algorithmic}
%\end{algorithm}

When comparing different variants, Askham and Kutz \cite{askham2018variable} suggest measuring the quality of a DMD method by comparing the reconstructed snapshots $\left( \bm{\Phi}(\bm{\lambda}) \bm{\Phi}(\bm{\lambda})^\dagger \mathbf{O}^\intercal \right)^\intercal$ with the original ones $\mathbf{O}$ such that:
\begin{equation}
	\text{Reconstruction quality} \equiv \frac{\| \mathbf{O}^\intercal - \bm{\Phi}(\bm{\lambda}) \bm{\Phi}(\bm{\lambda})^\dagger \mathbf{O}^\intercal \|_F}{\| \mathbf{O}^\intercal \|_F}
\end{equation}

%%=================================%
%----------- RESULTS --------------%
%==================================%
\section{Results}
\label{sec:results}
\subsection{Morphology and Classification}
\label{subsec:morphology}
\begin{figure}
	\centering
	\includegraphics[width=\textwidth]{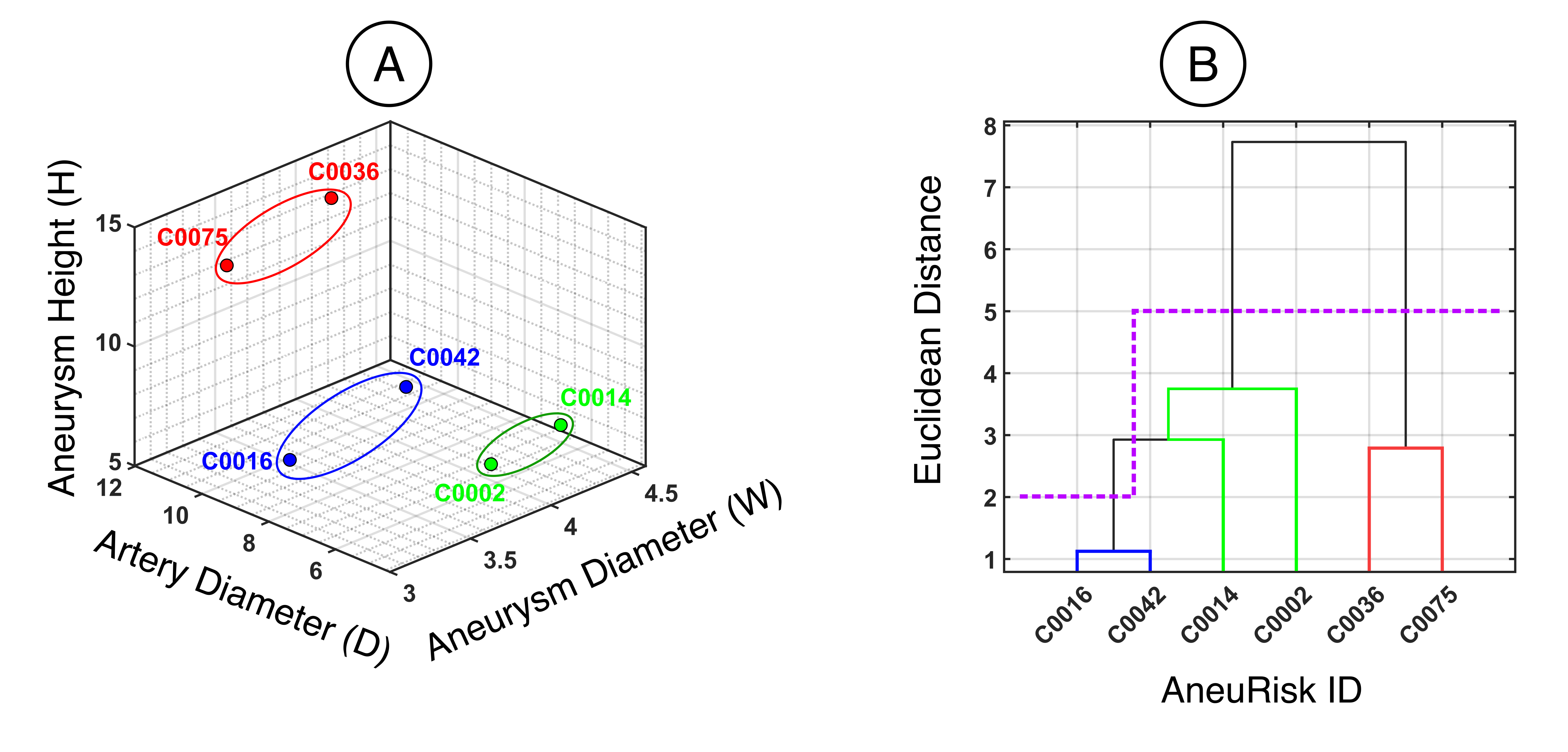}
	\caption{Patient-specific intracranial aneurysm (A) scattering and (B) hierarchical clustering based on morphological characteristics. The colored labels represent six representative patient IDs from the \textit{Aneurisk project}. Three different groups are consistent: small (green), medium (blue), and large (red) aneurysms. A dotted line represents the broken line for the clustering technique. The definitions of $D$, $W$, and $H$ are shown in Figure \ref{fig:setup}.}
	\label{fig:morphology}
\end{figure}

By qualitative analysis, three clusters can be identified from the relative locations of the patient-specific dataset in the 3D space depicted in Fig. \ref{fig:morphology}A. Patient 1 (ID: C0002) and Patient 2 (ID: C0014) are included in Group 1 (green cluster) situated at the bottom right, representing patients with small aneurysms. Group 2 (blue cluster) consists of Patient 3 (ID: C0042) and Patient 4 (ID: C0016) having medium-sized aneurysms. Group 3 (red cluster) consists of Patient 5 (ID: C0036) and Patient 6 (C0075), both are located high in the left corner which indicates their large size. 

The classification is quantified by hierarchical clustering, illustrated by the dendrogram in Fig. \ref{fig:morphology}B. Using the Euclidean distance at $2.0$ as the value to set the first broken line and working from bottom to top, the first clade joins the leaf of Patient 4 and the leaf of Patient 3. Hence, the two patients form a cluster (blue). Furthermore, the clade is located at $1.12$, indicating that the members of this cluster are the most similar. At the second clade ($2.92$) and the third clade ($3.74$), the right leaf always exhibits simplicifolious (single-leafed). Setting the next broken line at $5.0$ differentiates a cluster (red) on the right side. This cluster includes two individuals: Patient 5 and Patient 6. It is worth noting that the height difference between the two clades on the left side $\Delta h_{2-3} = 0.82$ is relatively small in comparison to $\Delta h_{1-2} = 1.71$ and $\Delta h_{3-4} = 3.98$. It shows that patients 1 and 2, while not in the same clade, are more similar to each other than they are to other clusters. Therefore, these two patients can form a cluster (green). Overall, both qualitative (scattered plot, see Fig. \ref{fig:morphology}A) and quantitative (hierarchical clustering, see Fig. \ref{fig:morphology}B) approaches divide our cohort into three distinct groups and arrive at the same members in each group.

The aspect ratio ($AR$) and the aneurysm number ($An$) are calculated for all the aneurysms and shown in Table \ref{tab:anatomical-parameters}. The values of the $An$, ranging from $2.7$ to $7.9$, are orderly arranged with the patient's numbering. On the contrary, there is no apparent correlation between the $AR$ and the group scheme identified in the previous analysis. The $AR$ in the study cohort varies from 0.75 to 1.80. The greatest $AR$ of 1.80 is exhibited by Patient 1, despite belonging to Group 1 (small aneurysms). Within this group, Patient 2's aneurysm has the lowest $AR$ of $0.75$, making it an opposite shape (short and wide) compared to the tall and narrow aneurysm of Patient 1. On the other hand, Patient 5 has a significantly smaller $AR$ of 1.32 while belongs to Group 3 (large aneurysms). The $AR$s of other patients (3, 4, and 6) are close to $1.0$ (blister shape). It is worth mentioning that patients 5 and 6 demonstrate a unique morphological characteristic, i.e., there are focal dilatations known as blebs (as noted in Table \ref{tab:anatomical-parameters}) developed on the wall inside the sac. Furthermore, both aneurysms have a distinct crescent shape in their sagittal footprints.

In Fig. \ref{fig:streamlines}, the CASes (represented by a black dash-dotted line) are used to align the 6 aneurysms in our cohort, and the flow direction is oriented from right to left. It can be observed that the patient cases in Group 1 possess obtuse aneurysm angles, forming by the CAS and the incoming flow (represented by a magenta dashed arrow). Specifically, this angle is very close to $90^{\circ}$ in the case of Patient 2. In contrast, the patient cases in Group 2 and Group 3 have more acute angles. It is important to mention that Patient 4 had the sharpest angle of inclination. In addition, the angle between the incoming flow and the existing flow (represented by an orange arrow) is qualitatively analyzed. The aneurysms in Group 1 are lateral because their angles exceed $90^{\circ}$. Meanwhile, individuals from the other two groups have terminal aneurysms.

\subsection{Patient-specific Hemodynamic Features from CFD Simulations}
\label{subsec:cfd}
There are two distinct hemodynamic features during the peak frame of the systole phase: (i) the inflow jet representing the large-scale dynamics, and (ii) the vortices representing the small-scale turbulence as shown in the instantaneous streamlines in Fig. \ref{fig:streamlines}. In this section, it is important to analyze the hemodynamics within the same group to highlight the relevance of aneurysmal morphology.

\subsubsection{Entrance of Inflow Jet}
\label{subsubsec:inflow_jet}
\begin{figure}
	\centering
	\includegraphics[width=\textwidth]{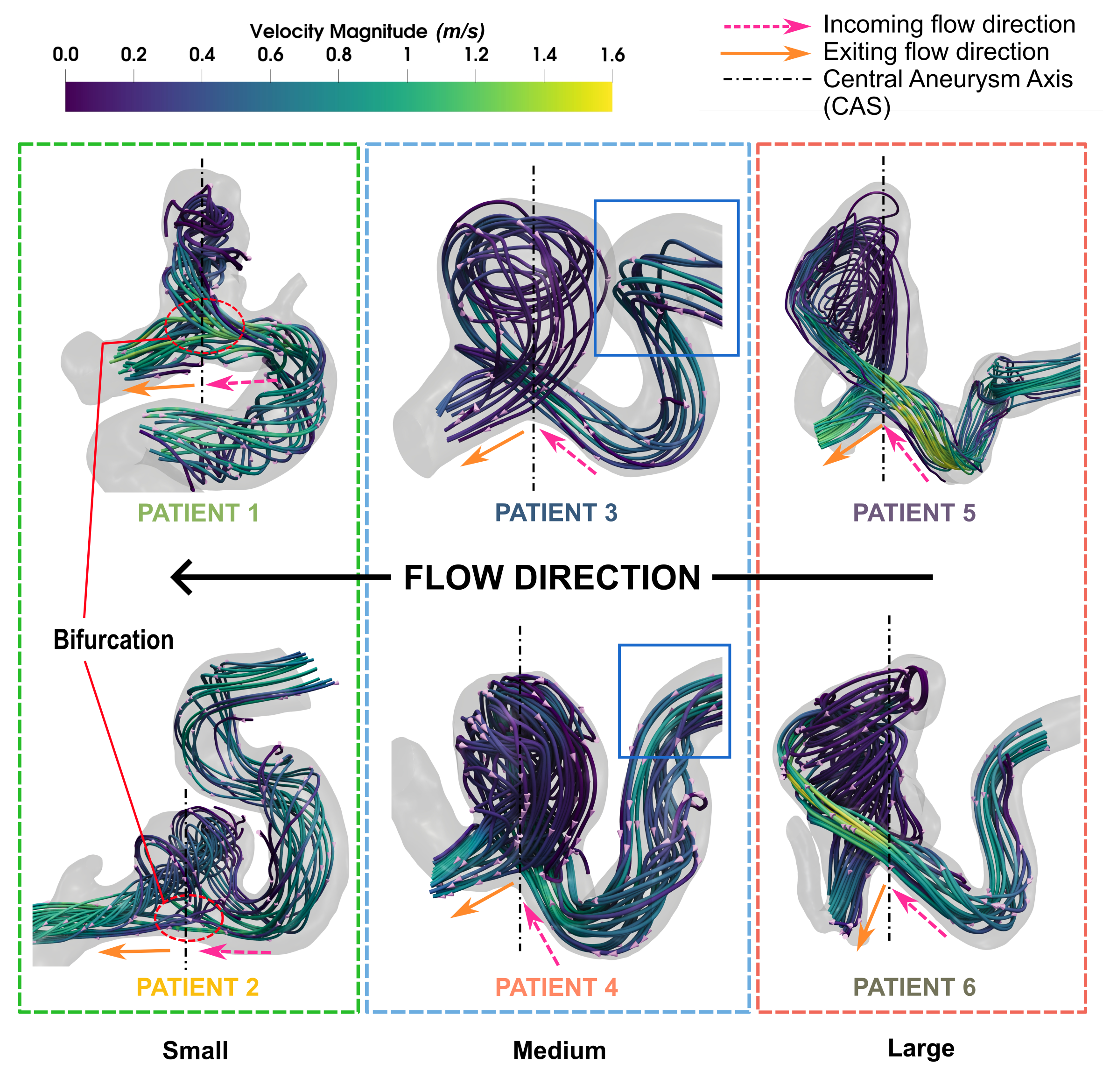}
	\caption{The streamlines are reconstructed from the velocity field at peak systole when the inflow jet is most prominent. The flow direction is from right to left. The aneurysms are aligned using the central aneurysms axes. In each case, a large vortex can be identified at the center of the aneurysm. Segments of the ICA are also included which exhibit patient specificity. The anatomical features of the aneurysms and the ICA dictate the incoming jet direction and influence the rotation of the central vortex.}
	\label{fig:streamlines}
\end{figure}

The cases in Group 1 show two distinct features in the dynamics of the inflow jet: (i) the splitting of the incoming flow; and (ii) the existence of helical flow in ICA. The splitting of the incoming flow is evident at the ostium, which is indicated by elliptically highlighted areas in Figure \ref{fig:streamlines}. Patient 2 receives a larger portion of the incoming flow, which demonstrates that the splitting feature can depend on the morphology (refer to the $AR$ as discussed in sec. \ref{subsec:morphology}) of the aneurysm. In both Patient 1 and Patient 2, the inflow jets enter the aneurysm sac at the posterior regions (note that aneurysm 1 is on the contralateral ICA). The inflow jet enters through the aneurysm ostium with a velocity of around $130$ cm/s and $60$ cm/s, respectively. The former velocity magnitude is much higher than the systolic peak of the bulk velocity at the inlet ($50$ cm/s). It is worth noting that the instantaneous streamlines in the cavernous (C4) and the C5 segments reveal a noticeable helical vortex in both cases.

In Group 2, the unique characteristic is the incoming flow traversing the aneurysm ostium near its central point. Depending on the incoming angle, the inflow jet can either approach the distal wall (Patient 3) or redirect toward the dome of the aneurysm (Patient 4). Because the morphologies of Patient 3 and Patient 4 are almost similar in terms of $AR$ (clustering results in Sec. \ref{subsec:morphology}), the bend curvature of the C4 segment of the ICA, indicated by squared highlighted areas in Figure \ref{fig:streamlines} for Patient 3, is responsible for the redirection. This demonstrates the influence of the incoming angle on the dynamics of the inflow jet. 

The patients in Group 3 exhibit the highest velocity in the inflow jet, evidenced by regions in Fig. \ref{fig:streamlines} having a magnitude exceeding $160$ cm/s. In the case of Patient 5, the high magnitude can be seen in the ICA segment before the incoming flow enters the aneurysm. Meanwhile, the inflow jet of Patient 6 achieves its highest magnitude after penetrating through the ostium. It is worth noting that the flow enters through the anterior region in this group. In addition, Group 3 possesses both differences in the aneurysmal and the ICA morphologies which have been observed separately in the previous groups as the contributor to the differences in the inflow jets' entrances.

\subsubsection{Impingement of Inflow Jet}
\label{sec:impingement}
Fig. \ref{fig:streamlines} shows that the inflow jets in patients 1, 2, 3, 5, and 6 are of the wall-impingement type. Meanwhile, Patient 4 represents the sole instance of the dome-impingement type within our cohort. It is noteworthy that within the population of wall-impingement occurrences, the locations of impingement are inconsistent. 

It is evident that the impingement type draws a bidirectional relationship between the characteristics of the inflow jet and the development of the aneurysms. Patient 4 (dome-impingement) has a pointy dome in comparison to a flatter one in the aneurysm of Patient 3 (wall-impingement). Again, analyses have shown that the two patients are very morphologically similar. More noticeably in patients 5 and 6 (wall-impingement), the jets directly hit the part of the distal wall where the middle section of the crescent aneurysms is located (see Sec. \ref{subsec:morphology}).

\subsubsection{Main Vortex Formation}
The main vortex is formed due to the separation of the inflow jet after making an impingement. To discern unique attributes in their formation and rotation, the relative angle between the centerline of the main vortex and the CAS is qualitatively examined in Fig. \ref{fig:streamlines}. In the case of Patient 1, after colliding with the distal wall in the lower chamber, the inflow jet disperses and continues in a spiral pattern toward the upper chamber. This spiral motion creates the main vortex rotating clockwise around the CAS of the aneurysm. Within our study cohort, Patient 6 is the only instance showing a similar swirling pattern around the CAS, even though the size of this patient's IA is two times larger (averaged to $12$ mm) than Patient 2's average size of $6$ mm. The main vortices in the cases of patients 2, 3, 4, and 5 are orientated perpendicular to their CAS, creating a contrasted appearance.

%In summary, this section shows that regarding the inflow jet, Patients 1 and 2 specifically exhibit bifurcation. Regarding impingement type, Patient 4 is the sole instance of dome-impingement. Regarding the main vortex, there exists a stark contrast in the vortical centerline and rotation. Here, the inconsistencies in the categorization based on the hemodynamic features of our cohort imply that aneurysm size is obsolete and one should look for an advanced way to quantify flow instabilities.

\subsection{Dynamic Mode Decomposition Results}
\label{subsec:dmd_results}
This section analyses results from the Hankel DMD method.

\subsubsection{The DMD Modes}
\label{subsubsec:dominant_modes}
\begin{figure}
	\centering
	\includegraphics[width=\textwidth]{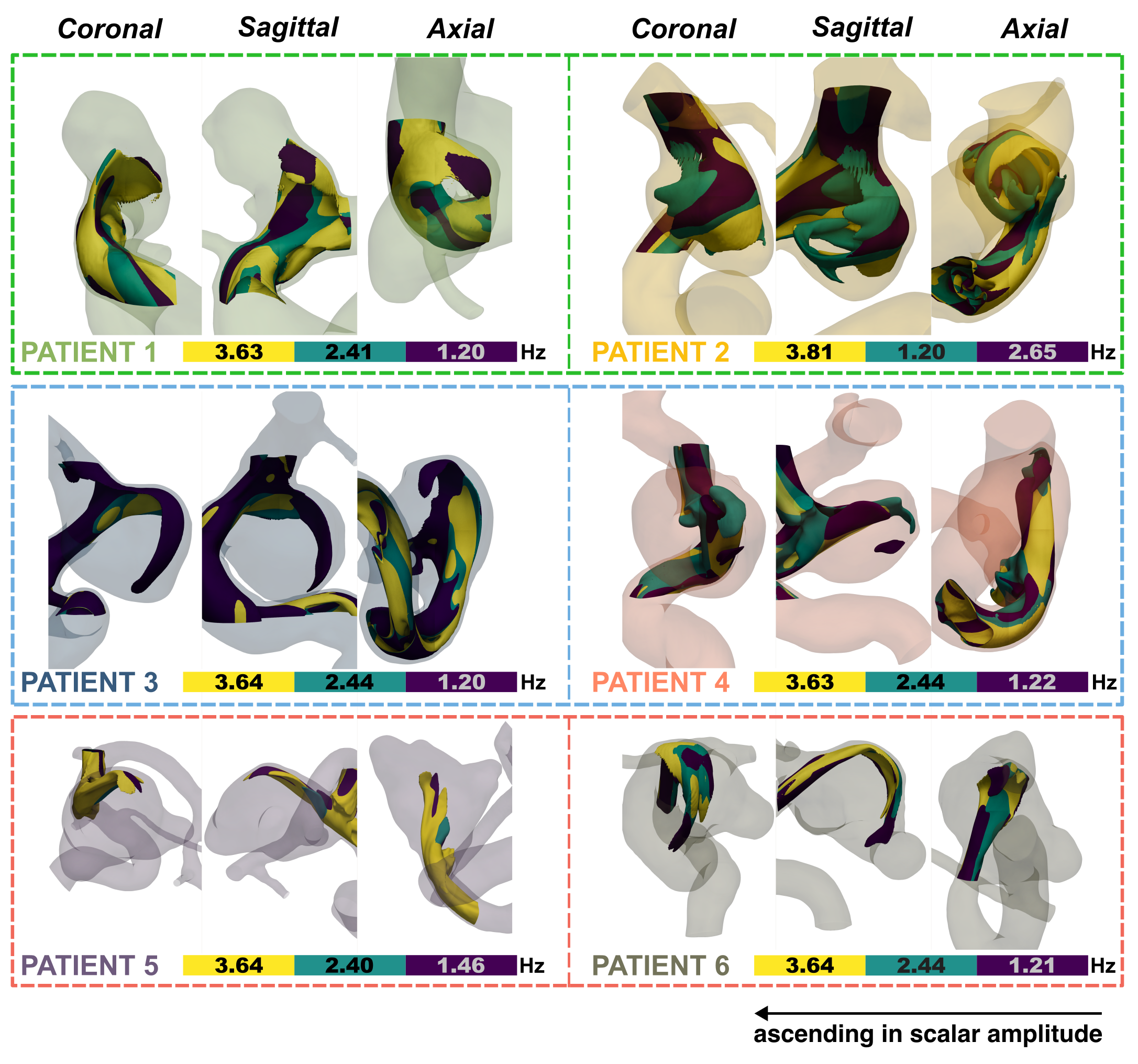}
	\caption{Three-dimensional structures of the three most energetic modes from the Hankel DMD method: i) Mode 1 (yellow); ii) Mode 2 (green); and iii) Mode 3 (purple). The order of the frequencies is related to their scalar magnitudes. The frequencies are close to the dominant frequencies (three Fourier modes) of the pulsatile flow at the inlet ($1.2$ Hz, $2.4$ Hz, and $3.6$ Hz). The iso-surface of the ratio $\frac{\mathbf{V}}{\mathbf{V}_{max}} = 0.5$ is used to visualize the structures. }
	\label{fig:dominant-modes}
\end{figure}

The Hankel DMD modes are sorted by their scalar amplitudes. Fig. \ref{fig:dominant-modes} shows the spatial structures (iso-contour) of the top 3 modes. The frequencies of these modes coincide with the frequency trio of the velocity waveform at the inlet (refer to Sec. \ref{subsec:comp_details}). In 5 out of 6 cases (patients 1, 3, 4, 5, and 6), the frequency $f_1 = 3.6$ Hz is consistently ranked as the most energetic mode (Mode 1 - yellow) followed by the energy of frequencies $f_2 = 2.4$ Hz and $f_3 = 1.2$ Hz as the second place (Mode 2 - dark green) and the third place (Mode 3 - purple), respectively. The case of Patient 2 displays a unique set of the dominant frequencies, i.e., $f_1 = 3.81$ Hz, $f_2 = 1.2$ Hz, and $f_3 = 2.65$ Hz, which differ slightly from those recorded in other patients.

The iso-contour of the three dominant modes effectively captures the visualization of the large-scale dynamics, the inflow jet. The spatial extents vary among the patient cases, as shown in Fig. \ref{fig:dominant-modes}. The traces of the inflow jets are immediately terminated after impingement in the cases of patients 1, 4, and 5. The inclusion of Patient 4 in this group indicates that the termination of the traces does not correlate with the impingement types specified in Section \ref{sec:impingement}. In contrast, the trajectory of the inflow jets of patients 3 and 6 can be captured after their impact with the distal walls, and the traces only terminate until they reach the domes. It is worth noting in most cases, the frequency of $1.2$ Hz possesses the highest spatial coverage and is often found in the terminal segment. The spatial extents in the case of Patient 2 possess uniqueness, again, from the study cohort, such that there exists a ring structure from Mode 2 ($f_2 = 1.2$ Hz).

In addition to the spatial extent of the iso-contour surfaces, the separation of the dominant modes can be qualitatively analyzed to classify the type of inflow jet. In this regard, patients 1, 2, and 3 exhibit "diffused" jets, while the jets are "concentrated" in the cases of patients 4, 5, and 6. One can see that the classification of the inflow jet does not correlate with the grouping scheme because both types of inflow jet are present in Group 2 (medium size). On the other hand, a comparison between two patients in Group 3 shows that two aneurysms in our cohort can belong to the same classification of inflow jet (concentrated) but differ in spatial extent.

\subsubsection{Pseudo-spectra}
\label{subsubsec:spectra}
The blood flow dynamics in IAs in one cardiac cycle are mostly contributed by the low-frequency ($< 15$ Hz) modes in all the cases. Fig. \ref{fig:selected_tempmag} shows the pseudo-spectra from 4 cases, i.e., Patient 1 (Group 1), Patients 3 and 4 (Group 2), and Patient 5 (Group 3), at the temporal resolution of $M = 200$ (represented by blue bars). The spectra increase uniformly from the beginning and peak at $3.6$ Hz. Observing the overall landscape, the top three modes are the ones captured in Sec. \ref{subsubsec:dominant_modes}. This quantity analysis in the temporal magnitude scale shows that the most energetic modes (ranked by scalar magnitude $| b_k |$) are also the most stable. After the peak at $3.6$ Hz, each spectrum exhibits a gradual decrease in normalized temporal magnitude. As noted from previous results (Fig. \ref{fig:dominant-modes}), the modes in the range $1-5$ Hz depict the inflow jet. Therefore, the dynamics in the upper range ($5-15$ Hz) are from the breakdown flows. The fact that the frequencies in this range are highly damped shows that these dynamics die out quickly in time.

The four cases in Fig. \ref{fig:selected_tempmag} share a common pattern, i.e., that the contributions of the DMD modes in the high-frequency domain ($> 15$ Hz) are small. None of the spectra sees a frequency peak above 40\% (represented by a red dashed line) of the maxima at $3.6$ Hz. It is worth mentioning that the landscapes exhibit patient specificity in both appearance and temporal magnitude. The spectra of patients 1 and 3 witness a flat landscape, with few frequencies barely reaching the $0.1$ line (represented by a green dashed line). However, Patient 3's landscape exhibits more fluctuation from $30$ Hz to $80$ Hz. Increasingly, the presence of fluctuation is not only more noticeable in the spectral landscape of Patient 4 but also appears instantly after $15$ Hz. High-frequency peaks around $20$ Hz, $65$ Hz and $110$ Hz reach up to 20\% magnitude of the maxima. Despite the two patients being the closest in terms of morphology, high-frequency is more pronounced in Patient 4, especially when comparing the domain from $65$ to $115$ Hz. Concerning Patient 5, the spectrum is flat from $15-110$ Hz, a rather stark contrast to the smaller aneurysms in Group 2. There exists one peak of frequency at $117.54$ Hz accounting for a temporal magnitude equivalent to 25\% of the maxima.

The contribution of high-frequency fluctuations can be great in certain aneurysms, those are the cases of patients 2 and 6. Fig. \ref{fig:peak_modes}A depicts their spectra showing that there exist domains in the high-frequency region that are as stable as the large-scale dynamics. In the case of Patient 2, the majority of the DMD modes before $110$ Hz surpass the 10\% line. One can clearly notice that the domain from $40$ to $70$ Hz stands out because it includes frequencies passing the 40\% line (represented by a red dashed line). The landscape in this domain includes two peaks at $50.90$ Hz and $60.70$ Hz accounting for normalized magnitudes of 0.96 and 0.80, respectively. It is worth noting that Patient 2's spectrum witnesses significant fluctuations around $10$ Hz, which is yet another uniqueness of this patient. In the case of Patient 6, the active domain is at higher frequencies ($>70$ Hz). This shift creates a frequency gap between the large-scale dynamics and the high-frequency fluctuations, in which the active domain of Patient 2 ($40-70$ Hz) sits in between. In contrast to the previous case, these frequencies in the spectrum of Patient 6 are damped. The active domain in this case begins at $72.85$ Hz and ends after $82.43$ Hz. It experiences more volatility and sees only one substantial peak at $76.93$ Hz, which has a temporal magnitude accounting for 80\% of the maxima.

\subsubsection{Location of Significant High-Frequency Fluctuations}
\label{subsubsec:high_freq}
The iso-contours of the DMD mode in interested domains are reconstructions for Patients 2 and 6. In Fig. \ref{fig:peak_modes}B, the spatial structures of these modes appear fractured, which is in juxtaposition to the smooth and continuous surfaces reconstructed using the dominant modes (refer to Fig. \ref{fig:dominant-modes}). The spatial extents of the high-frequency modes are located where the inflow jets (from the dominant modes, depicted in gray) separate after the impingements and form vortical structures in the aneurysm sac. In Patient 2, the fragments of high-frequency modes concentrate at the sac's volume center at which the main vortex is formed. Meanwhile, in Patient 6, the high-frequency modes are evident after the inflow jet impinges on the aneurysm's distal wall.

\subsubsection{Impact of Temporal Resolution on Pseudo-spectra}
\label{subsubsec:low_temp_res_spectra}
The frequency bandwidth scales proportionally with the temporal resolution. At $M = 50$, all spectra are maxed out at $30$ Hz. In Fig. \ref{fig:selected_tempmag} and Fig. \ref{fig:peak_modes}A, the landscapes of the spectra for $M = 50$ (represented by yellow bars) bear resemblance to their corresponding sections at $M = 200$ (highlighted by grey zones). By quantitative assessment, the frequencies are identical between the two resolutions. In addition, the similarity is qualitatively higher in the group of 4 patients having damped high-frequency modes. A consistent pattern emerges across the cohort, i.e., frequencies ranging from $5$ to $15$ Hz gain higher temporal magnitudes and frequencies beyond $20$ Hz show the opposite pattern. Ultimately at this reduced temporal resolution, high-frequency fluctuations are absent.

\subsection{Cumulative Energy Curves from DMD Modes}
\label{subsec:ce_curves}
The efficacy of the Hankel DMD method is quantitatively assessed in this section. Under various spatiotemporal resolutions, the normalized cumulative energy (CE) curves are constructed from the values $\sigma_i$ of the singular matrix $\mathbf{\Sigma}$. The DMD data from 72 standard patient-specific cases (refer to Section \ref{sec:scope}) is carried out by employing a sensitivity analysis with respect to spatial resolution and temporal resolution. In addition, the choice VOI is also considered an independent variable. Each section below investigates the influence of one while keeping all other parameters constant.

\subsubsection{Pattern of Patients' CE Curves}
\label{subsubsec:stratification}
In Fig. \ref{fig:all-temporal}A, the patients' normalized CE curves appear to have no overlaps at $M = 50$. The curves are separated and their order can be identified with ease especially when scaling the CE plot in the log-log scale. Here, the order of the patients from top to bottom is as follows: Patient 1 $\rightarrow$ Patient 3 $\rightarrow$ Patient 6 $\rightarrow$ Patient 4 $\rightarrow$ Patient 5 $\rightarrow$ Patient 2. It is worth noting that a significant gap is present between patients 3 and 6, consequently, creating a visual barrier that divides our cohort into 2 groups: i) the upper, and ii) the lower. The gap gains substantial evidence when the curves reach higher CE levels.

To further quantify the distance between the two groups, a vertical line (dotted cyan) is taken as a reference line at Mode 10 (which is equivalent to 20\% of the total number of modes). CE value at the reference line is measured for each case, and then the average value can be calculated from the members of the same group. The upper group consists the case $P_1-1\mathsf{X}-M_{50}$ (0.91) and the case $P_3-1\mathsf{X}-M_{50}$ (0.88) which averages to 0.89. In the lower group, the cases $P_2-1\mathsf{X}-M_{50}$, $P_4-1\mathsf{X}-M_{50}$, $P_5-1\mathsf{X}-M_{50}$, and $P_6-1\mathsf{X}-M_{50}$ reach, respectively, 0.78, 0.81, 0.77, and 0.82 in normalized energy after 10 modes. This group averages 0.79 resulting in a normalized energy gap of 0.1 with the upper group, or 10\% of the total energy.

Taking advantage of the log scale, another quantitative approach is used employing linear fitting. The technique calculates the best-fit lines using the CE values of the first 10 modes for each group. Fig. \ref{fig:all-temporal} shows that our approach yields two optimal lines that differ in slope. The optimal line for the upper group (in blue) has a slope of 0.27. Whereas, the slope for the lower group's line is 0.22. This analysis presents an index showing that the upper cases have fast decay curves, and the ones from the lower cases exhibit slower decay.

\subsubsection{Single Effect of Temporal Resolution}
\label{subsubsec:temp_res_ce_curves}
In this sensitivity study, the Hankel DMD method is applied to the patient-specific dataset at the original resolution. A combination of 6 aneurysms and 3 temporal resolutions in this section yields 18 cases with the following notations $P_1-1\mathsf{X}-M_{50}$, $P_2-1\mathsf{X}-M_{50}$, $P_3-1\mathsf{X}-M_{50}$, $P_4-1\mathsf{X}-M_{50}$, $P_5-1\mathsf{X}-M_{50}$, $P_6-1\mathsf{X}-M_{50}$, $P_1-1\mathsf{X}-M_{100}$, $P_2-1\mathsf{X}-M_{100}$, $P_3-1\mathsf{X}-M_{100}$, $P_4-1\mathsf{X}-M_{100}$, $P_5-1\mathsf{X}-M_{100}$, $P_6-1\mathsf{X}-M_{100}$, $P_1-1\mathsf{X}-M_{200}$, $P_2-1\mathsf{X}-M_{200}$, $P_3-1\mathsf{X}-M_{200}$, $P_4-1\mathsf{X}-M_{200}$, $P_5-1\mathsf{X}-M_{200}$, $P_6-1\mathsf{X}-M_{200}$.

The pattern of the two groups identified in Sec. \ref{subsubsec:stratification} remains unchanged when the temporal resolution increases to $M = 100$ and $M = 200$. Increasingly, the gap between the groups is consistent at around 10\%. In Fig. \ref{fig:all-temporal}B, taking the same measurements and calculations at Mode 20 (20\% of the total 100 modes) shows that the upper and lower averages, respectively, are 0.94 and 0.84. In Fig. \ref{fig:all-temporal}C, the averaged values at Mode 40 (20\% of the total 200 modes) are 0.98 for the upper group and 0.89 for the lower group. In addition, comparing the optimal lines in Fig. \ref{fig:all-temporal}A, B, and C shows that the slopes decrease when more snapshots are added to the DMD input. Specifically, the upper group exhibits a rate of 0.01. The lower group decreases at a rate that is twice as fast showing a rate of 0.02.

\subsubsection{Single Effect of Spatial Resolution} 
\label{subsubsec:spat_res_ce_curves}
Next, the temporal resolution is kept constant at $M = 200$ while the spatial resolution varies. The analysis in this section presents two patients, and they are the representatives of their groups (refer to Fig. \ref{fig:all-temporal}), i.e., Patient 1 is in the fast-decay group and Patient 6 is in the slow-decay group. As Patient 6 has been shown to exhibit significant high-frequency fluctuations (refer to Fig. \ref{fig:peak_modes}B), we perform an addition resampling case for this patient's CFD data having a coarsening factor of $16\mathsf{X}$ to achieve a low spatial resolution ($> 1.0$ mm) normally obtained from 4D-flow MRI measurements. The 8 cases are denoted as $P_1-1\mathsf{X}-M_{200}$, $P_1-2\mathsf{X}-M_{200}$, $P_1-4\mathsf{X}-M_{200}$, $P_1-8\mathsf{X}-M_{200}$, $P_6-1\mathsf{X}-M_{200}$, $P_6-4\mathsf{X}-M_{200}$, $P_6-8\mathsf{X}-M_{200}$, $P_6-16\mathsf{X}-M_{200}$. In addition, the CE values in the cases with original resolution ($P_1-1\mathsf{X}-M_{200}$ and $P_6-1\mathsf{X}-M_{200}$) are chosen to be the baselines. Consequently, error bounds are calculated using the baselines.

\begin{figure}
	\centering
	\includegraphics[width=\textwidth]{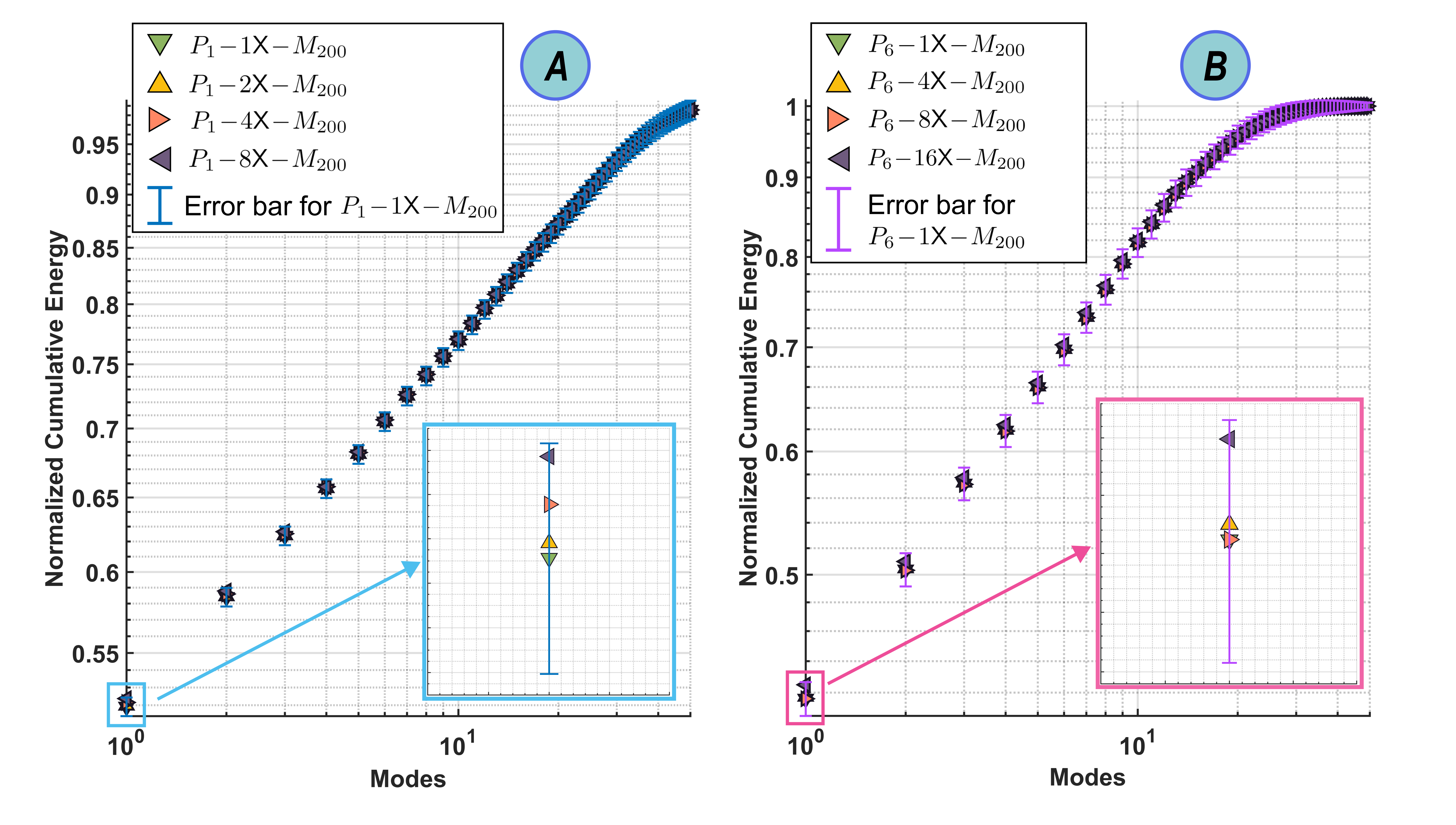}
	\caption{The consistency of the cumulative energy (CE) curves across all spatial resolutions $1X, 2X, 4X, 8X$ (see Table \ref{tab:spatial-resolution}) at fixed temporal resolution $M = 200$. The RICs are plotted for (A) Patient 1 and (B) Patient 6 in a log-log scale. The differences among the normalized CE curves are represented with the error bounds of $\pm 1.0\%$ (cyan) for Patient 1 and $\pm 1.5 - \pm 0.5$ (magenta) in Patient 6. The discrepancies among the CE values for all patients are less than $\pm 1.5\%$ error. After $20\%$ of mode, the errors diminish. The insets (cyan and magenta) illustrate the magnified views in the first mode.}
	\label{fig:spatial}
\end{figure}

The overall disparities in the CE curves in Fig. \ref{fig:spatial}A and Fig. \ref{fig:spatial}B among different spatial resolutions from $2\mathsf{X}$ to $8\mathsf{X}$ fall within 1\% error from the baselines for both cases of patients 1 and 6. The biggest error between any low-resolution curve and the baseline is in the first DMD mode. By examining the magnified views of this mode included in Fig. \ref{fig:spatial}, it becomes apparent that the lower-resolution curves exhibit an up-shifting behavior. However, this behavior is indeterminable. In the case of Patient 1, the increments in the energy of the first mode are present (see the magnified view in Fig. \ref{fig:spatial}A). However, the increment is not linearly proportional to the coarsening factor. For example, the first mode of the normalized CE curve from $P_6-8\mathsf{X}-M_{200}$ is closer to the baseline than the first mode from $P_6-4\mathsf{X}-M_{200}$ (see the magnified view in Fig. \ref{fig:spatial}B). Nevertheless, at an extremely coarse data of $P_6-16\mathsf{X}-M_{200}$, the disparity at the first mode is less than 1.5\% error from the baseline. As the CE curves progress, the errors get smaller as values from cases at different temporal resolutions converge.

\subsubsection{Single Effect of Volume of Interest}
\label{subsubsec:voi_choice_ce_curves}
Finally, this section analyses the influence of various scenarios when extracting the VOI. Using a similar approach, the temporal resolution is fixed at $M = 200$, and the CFD data is kept at the original resolution. The cases are notated as $P_4-1\mathsf{X}-M_{200}$ to reflect the current settings. 

\begin{figure}
	\centering
	\includegraphics[width=\textwidth]{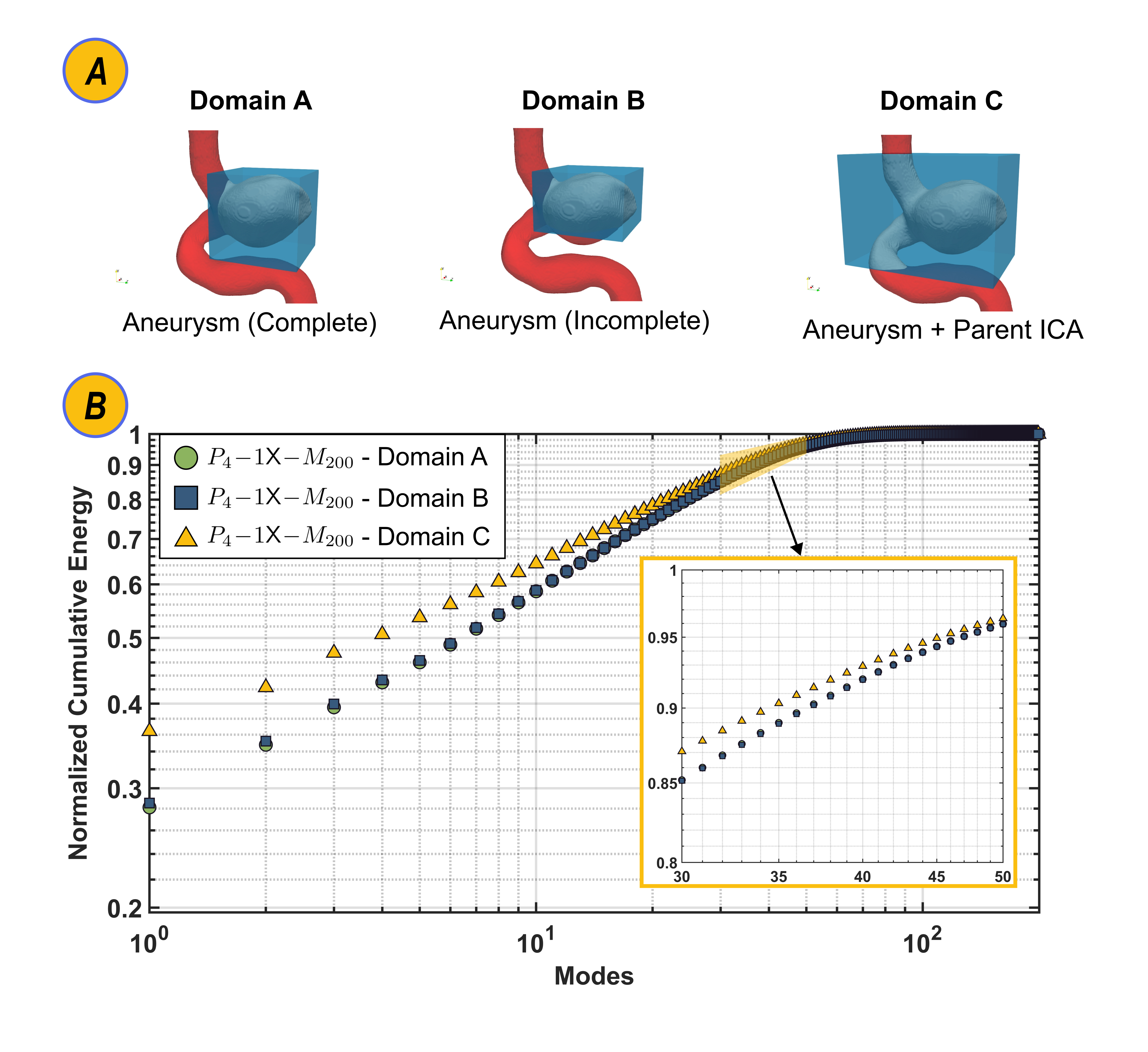}
	\caption{The impact of volume of interest (VOI) on DMD analysis in Patient 4 ($1\mathsf{X}$) for $M = 200$: (A) Complete aneurysm sac; (B) Incomplete sac; and (C) Aneurysm sac and parts of the ICA. The CE curves are consistent for cases (A) and (B). However, including the parent artery strongly interferes with the results of DMD analysis. The curves are shown in a log-log scale.}
	\label{fig:domain-impact}
\end{figure}

Fig. \ref{fig:domain-impact}A illustrates three possible options for the VOI of Patient 4. This patient is chosen out of six individuals in the cohort because they represent the intricacy of aneurysm location that influences the extracting process. Specifically, the patient's aneurysm developed in the vicinity of the bend from the C4 to the C6 segments (refer to Fig. \ref{fig:setup}A). Therefore, a section of the internal carotid artery (ICA) can interfere with the VOI box. Three distinct VOIs have been established for Patient 4 as follows: i) Domain A encompasses the complete aneurysm sac, a portion of the ICA's C4 in the caudal volume, and the posterior boundary truncated at the ostium; ii) Domain B comprises an incomplete sac by omitting the protrusion of the proximal wall in the caudal volume, this avoids the C4's inclusion; iii) Domain C includes the complete sac along with arteries of the ICA in the posterior direction. 

The absence of aneurysmal volume at the proximal wall (Domain B) produces a minimal influence on the normalized CE curve. Still, the curve slightly shifts upward when compared to the curve obtained from Domain A (baseline) in Fig. \ref{fig:domain-impact}B. The behavior of the CE curve in the case of Domain B can be seen as similar to the cases with low spatial resolutions. Fig. \ref{fig:domain-impact}B shows that there is a significant displacement in the normalized CE curves obtained from Domain C compared to the baseline. In detail, adding sections of the parent artery raises the energy of the first mode by a staggering 28\% from the baseline value. The inset in Fig. \ref{fig:domain-impact}B as a magnified view of the region of modes $30-50$ shows that after mode 40, the values from the three curves have less than 1\% error. Consequently, the normalized CE curve from Domain C exhibits the most significant deviation.

\subsection{Performance of Hankel DMD and Optimized DMD}
\label{subsec:2dmd}
\subsubsection{Reconstruction Quality}
\label{subsec:2dmd_quality}
\begin{figure}
	\centering
	\includegraphics[width=\textwidth]{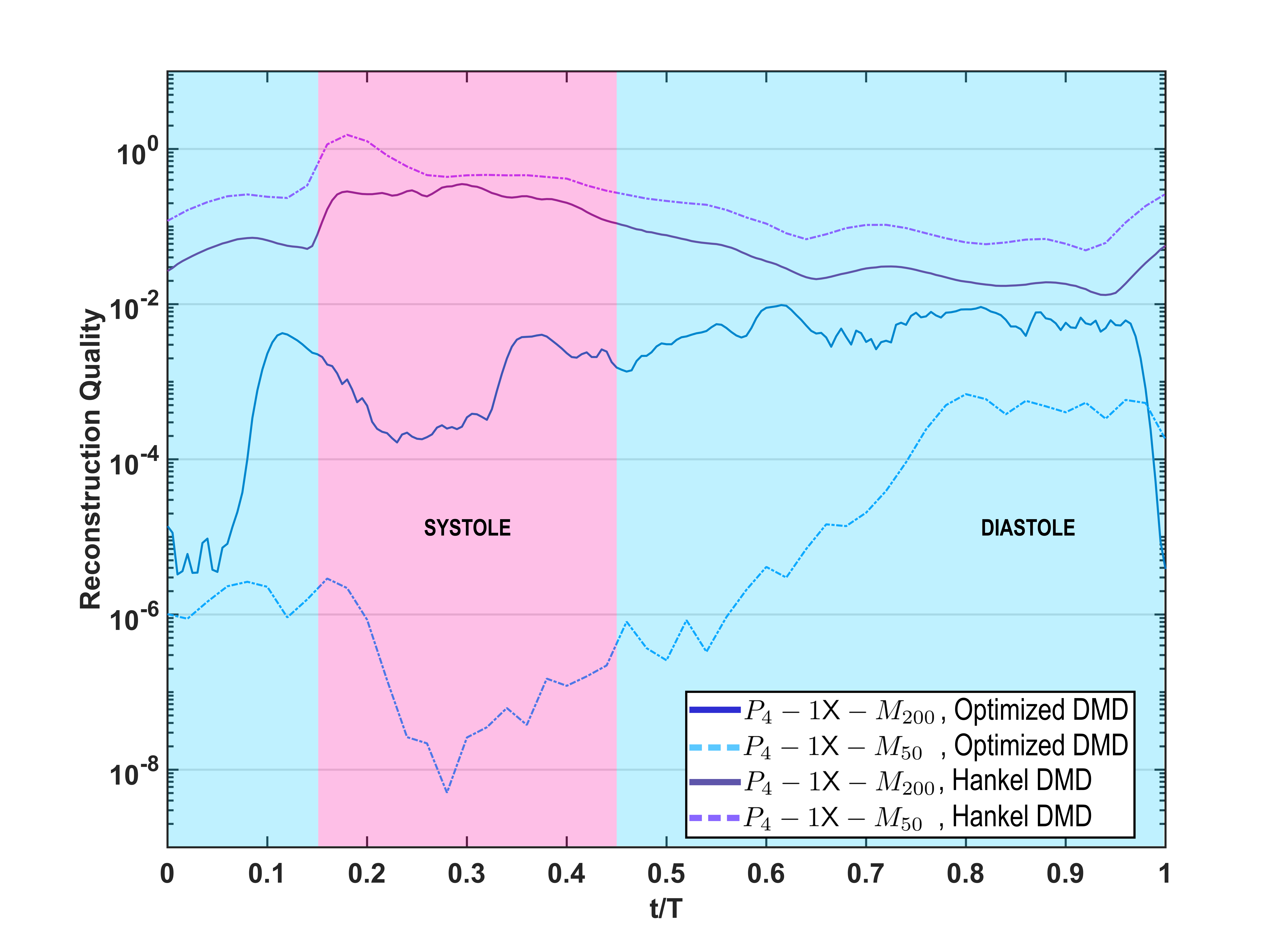}
	\caption{Performance comparison between Hankel and Optimized DMD analysis. The Frobenius norm error of the reconstructed snapshots (Patient 4) over one cardiac cycle shows that the Optimized DMD method has better reconstruction quality (lower errors) at both low ($M=50$) and high ($M=200$) temporal resolutions.}
	\label{fig:reconstruction_quality}
\end{figure}

The Optimized DMD method outperforms the Hankel method in terms of dynamic accuracy because it has lower reconstruction error, especially during the systole period for both cases $P_4-1\mathsf{X}-M_{50}$ and $P_4-1\mathsf{X}-M_{200}$ as shown in Fig. \ref{fig:reconstruction_quality}. The Optimized DMD method achieves remarkably high precision during the systole phase (error $\approx 1 \times 10^{-8}$). As the temporal resolution increases to $M = 200$, the quality curve of Optimized DMD (in blue) attains significantly smaller errors ($\approx 1 \times 10^{-6}$) compared to the curve from the Hankel method (in purple). There exists a stark contrast as the average error for Hankel DMD is $\approx 1 \times 10^{-1}$, with notably high errors during systole (from $t/T = 0.15$ to $t/T = 0.45$). In addition, the diastole phase at $M = 200$ sees the errors between the two methods become close to each other in the range $1 \times 10^{-2}$. Ultimately, the Optimized DMD still outperforms Hankel DMD in terms of reconstruction quality during diastole, even at their peak error at around $t/T = 0.6$, regardless of temporal resolution.

The Hankel DMD maintains a consistent quality curve shape across resolutions, while the opposite is true for the Optimized DMD method. Increasingly, the reconstruction quality of the Hankel DMD method improves as temporal resolution increases. In contrast, the increase in temporal resolution worsens the performance of Optimized DMD. The evidence is clear when comparing the value of the cases $P_4-1\mathsf{X}-M_{50}$ and $P_4-1\mathsf{X}-M_{200}$ from Hankel DMD at any timestamp in Fig. \ref{fig:reconstruction_quality}. Both lines remain at low error, then sharply increase during early systole. The peaks are at around $t/T = 0.17$, and then the quality curves gradually decrease throughout the late systole and diastole phases. The result from the Optimized method shows a partially similar trend at the temporal resolution $M = 200$, but there are differences. First, the most noticeable is the skyrocketed period when entering the systole phase. However, the peak can be seen at an earlier time $t/T = 0.13$. Second, the Optimized DMD method's precision decreases with time in the cardiac cycle, with a significant error increase (two orders of magnitude) in systole, becoming especially clear as $t/T > 0.65$ (diastole). Interestingly, with $M = 50$, the increasing error theme during the systole phase in the quality curve of the Optimized method is replaced by a decreasing one. The performance of the Optimized DMD across different temporal resolutions, as evidenced in Fig. \ref{fig:reconstruction_quality}, exhibits an unpredictable nature in maintaining consistent reconstruction quality.

\subsubsection{Decomposed Dynamic Modes}
\label{subsec:2dmd_modes}
\begin{figure}
	\centering
	\includegraphics[width=\textwidth]{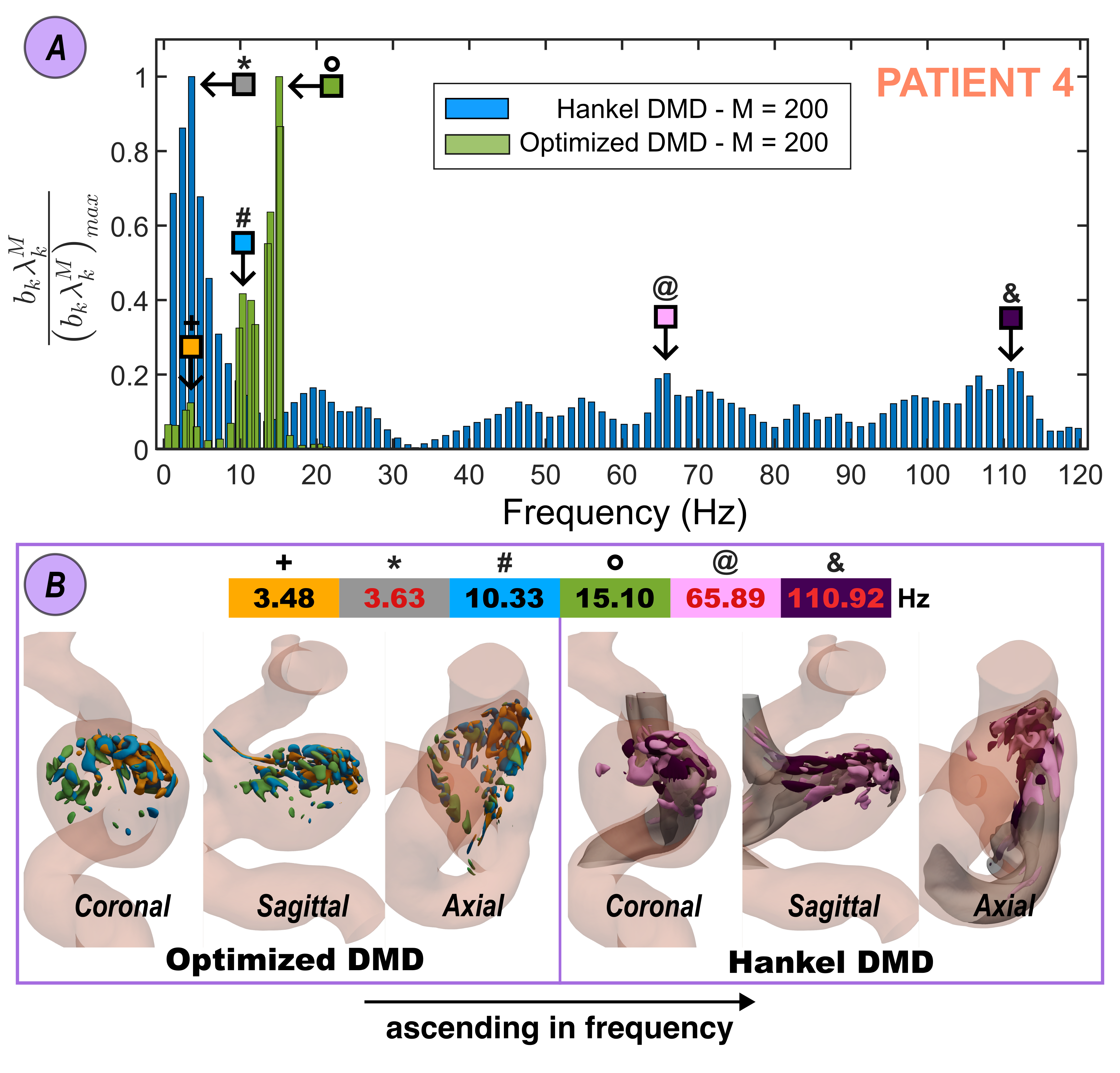}
	\caption{Distributions of Hankel and Optimized DMD modes in Patient 4 at $M = 200$. (A) The dominant modes of Optimized DMD concentrate in a lower range of frequencies ($< 15$ Hz). (B) The 3D structures of both Optimized and Hankel DMD modes indicate the interaction with the distal wall.}
	\label{fig:patient4_optDMD}
\end{figure}

The above results show that the differences in the algorithm of the two methods influence the identification of hemodynamic features. To delve deeper into the analysis between Hankel and Optimized DMD methods, the pseudo-spectra of Patient 4's dynamic modes is depicted in Fig. \ref{fig:patient4_optDMD}(A). A clear distinction can be seen between Hankel and Optimized DMD methods. In Hankel DMD, the dominant frequency ($f_2 = 3.63$ Hz) agrees well with ones from the inflow waveform (refer to Fig. \ref{fig:setup}). The highest frequencies are sustained from $65.89$ Hz to $110.92$ Hz. However, the Optimized DMD does not provide such a frequency range. Instead, the spectrum of Optimized DMD concentrates to a much lower range of $20$ Hz with a peak of $15.1$ Hz. In other words, the Optimized DMD does not provide the expected frequencies of the inflow waveform. It also misses all frequencies above $16$ Hz, which indicates that the high-frequency modes of Optimized DMD have a minimal contribution to the reconstruction of large-scale hemodynamics.  

The spatial extents of the two methods' DMD modes are depicted in Fig. \ref{fig:patient4_optDMD}(B). The Optimized DMD provides the dominant modes at different frequencies $f_1 = 3.48$ Hz, $f_2 = 10.33$ Hz, and $f_3 = 15.10$ Hz (see Fig. \ref{fig:patient4_optDMD}). Regarding the dominant frequency of $15.10$ Hz, its fragments were scattered densely near the proximal wall. Hankel DMD provides a well-defined structure for the inflow jet (shadowed - $1.2$ Hz, $2.4$ Hz, and $3.6$ Hz). The higher frequencies ($15.1$ Hz, $65.89$ Hz, and $11.92$ Hz) localize around the inflow jet. The location of frequency $3.48$ Hz coincides with the location of impingement. In comparison to the peak frequencies of the Hankel DMD approach, the accuracy to identify the location of impingement was significantly higher using the Hankel dominant mode (3.63 Hz). In addition, the traces of high frequencies from $65$ Hz to $115$ Hz concentrated around the inflow jet indicated the breakdowns of the jet at impact and its interactions with the vortices.

\subsubsection{Eigenvalues}
\label{subsec:2dmd_eigenvalues}
\begin{figure}
	\centering
	\includegraphics[width=\textwidth]{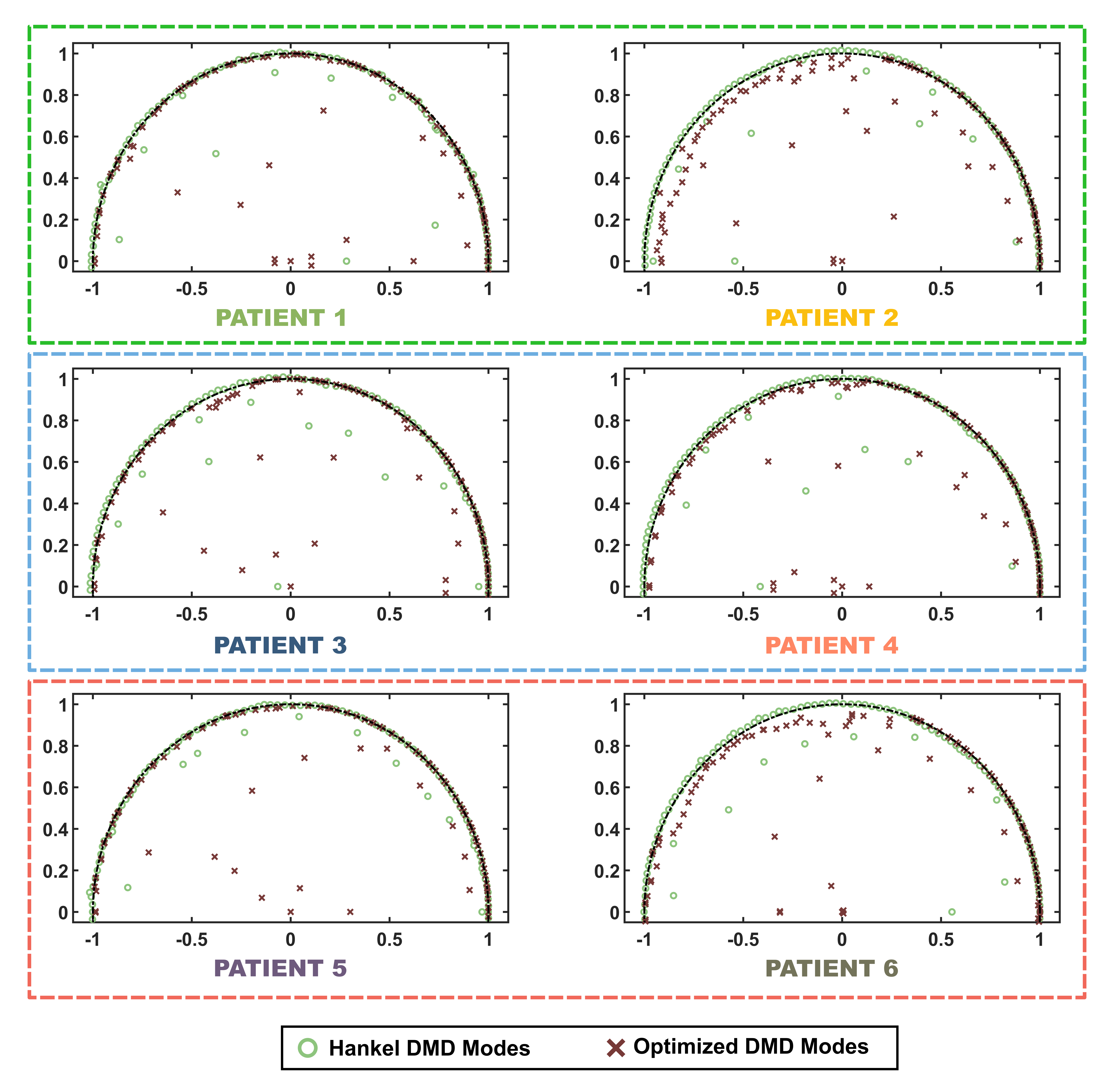}
	\caption{Distribution of eigenvalues ($\lambda_i$) on the unit circle from the Hankel and Optimized DMD algorithms for all patients. The distribution of $\lambda$ indicates a significant number of transient dynamics over the cardiac cycle. The Hankel eigenvalues (circle symbol) are significantly more stable than the ones from the Optimized method (cross symbol). The symbols are non-overlapping showing that the frequencies are different in the two methods. Due to the symmetry of a $\lambda$ pair, only the upper half of the unit circle is shown. Aneurysms are categorized by their sizes into small (Group 1 - green), medium (Group 2 - blue), and large (Group 3 - red).}
	\label{fig:unitCircle}
\end{figure}

Fig. \ref{fig:unitCircle} demonstrates that Hankel DMD modes exhibit greater stability compared to Optimized DMD modes throughout the cardiac cycle across all six patients. It is most evident when comparing the eigenvalue spectra in six cases $P_1-1\mathsf{X}-M_{200}$, $P_2-1\mathsf{X}-M_{200}$, $P_3-1\mathsf{X}-M_{200}$, $P_4-1\mathsf{X}-M_{200}$, $P_5-1\mathsf{X}-M_{200}$ and $P_6-1\mathsf{X}-M_{200}$. More Hankel eigenvalues localize on the unit circle's circumference than Optimized eigenvalues. Most Optimized DMD eigenvalues within the left semicircle are contained inside the circle, indicating damping in the corresponding dynamic modes. Consequently, stable modes from the Optimized DMD method concentrate on the right side of the circumference (first and fourth quadrants), corresponding to frequencies below 16 Hz with significantly higher contributions (Fig. \ref{fig:patient4_optDMD}A). The disparity in eigenvalue distribution between the two methods reflects their different frequency dispersions.

In conclusion, while the Optimized DMD method shows superior reconstruction quality, especially at lower temporal resolutions, the Hankel DMD method demonstrates better frequency range coverage, more accurate representation of flow features, and greater eigenvalue stability.

%%=================================%
%---------- DISCUSSION ------------%
%==================================%
\section{Discussion}
\label{sec:discussion}
\subsection{Morphological Indices and the Complexity of Aneurysmal Flows}
The complexity of flow patterns within saccular aneurysms has long been postulated to correlate with morphological characteristics such as size or aspect ratio ($AR$), which can be readily measured using contemporary imaging technologies like MRI or CT \cite{dhar2008morphology, ujiie2001aspect, wuetal2021hemo}. While aneurysms in Group 1 are considered small in all evaluation techniques, all aneurysms in Group 2 and Group 3 surpass the critical size threshold of stability for both PHASES system ($7.0$ mm) and the ELAPSS system ($5.0$ mm) \cite{BRINJIKJI2018e425}. Traditionally, an $AR$ exceeding $1.6$ is recognized as a potential indicator of increased rupture risk \cite{ujiie2001aspect, nader2004aspect}. This association draws on analogies with fluid dynamics in a lid-driven cavity \cite{zhang.2022}, where higher $ARs$ are thought to render complex flow patterns with multiple vortices. However, our results in Fig. \ref{fig:streamlines} present a more nuanced view as discussed below.
\begin{itemize}
    \item[a] \textbf{\textit{Aspect Ratio}}: Within our study cohort, only Patient 1 exhibits an $AR$ greater than $1.6$ (see Table \ref{tab:anatomical-parameters}) but its flow dynamics is rather similar to other cases as shown in Fig. \ref{fig:streamlines}. A careful examination of results in Figures \ref{fig:dominant-modes}, \ref{fig:selected_tempmag}, and \ref{fig:all-temporal} shows that the flow within Patient 1’s aneurysm is dominated by large-scale flow structures with flow-frequency fluctuations ($< 15$ Hz). On the contrary, the $AR = 0.75$ of Patient 2's aneurysm (see Table \ref{tab:anatomical-parameters}) results in more complex flow patterns with significant contribution of high-frequency dynamics, especially around $60.70$ Hz (Fig. \ref{fig:selected_tempmag}). In addition, the CE curve of Patient 2 consistently has the lowest decay rate in Figures \ref{fig:all-temporal}A, \ref{fig:all-temporal}B, and \ref{fig:all-temporal}C, which indicates the presence of transition-to-turbulence. As Patient 1 and Patient 2 belong to the same Group 1 (small aneurysms), these observations suggest that $AR$, in isolation, is insufficient in predicting the intricacies of flow within aneurysms. This discrepancy underscores the limitations of using $AR$ as a standalone predictor of flow complexity and emphasizes the importance of additional geometric and hemodynamic factors.
    \item[b] \textbf{\textit{Aneurysm Number}}: $An$ \cite{le2010pulsatile} was introduced to alleviate the shortcomings of morphological indices such as $AR$ in predicting aneurysm flow dynamics by incorporating the flow waveform, i.e., $PI$. Our data, as shown in Table \ref{tab:anatomical-parameters} and Fig. \ref{fig:peak_modes}, indicates that both the Aneurysm Number ($An$) and the Width-to-Depth ($W/D$) ratio exceed 1 for all patients. According to previous studies \cite{le2010pulsatile, le2013vortex}, these matrices are indicative of potential high-frequency fluctuations \cite{asgharzadeh2020simple}. However, a direct correlation between $An$ and the peak frequencies of these fluctuations cannot be established in the current patient cohort. For example, a distinct peak frequency of $117.5$ Hz is noted in Patient 5 with an $An$ of $6.3$, while frequencies around $75.0$ Hz are observed in Patient 6 ($An = 7.9$), as detailed in Fig. \ref{fig:selected_tempmag}. This variability suggests that while $An$ and $W/D$ ratios can signify the presence of complex flows, they do not consistently predict the exact frequencies of flow fluctuations. Therefore, the reliance on $An$ for predicting aneurysm rupture risk should be approached with caution \cite{asgharzadeh2020simple, yang2024verification}. 
    \item[c] \textbf{\textit{Aneurysmal angles}}: It is crucial to consider additional geometric factors, such as the flow angle, which can significantly influence flow dynamics as depicted in Fig. \ref{fig:dominant-modes}. This multifactorial approach might be essential for a more accurate assessment of rupture risk and understanding of aneurysm growth.
\end{itemize}

\subsection{Inflow Jet Dynamics}
\textit{In-vivo} flow measurements suggest that the inflow jet (which induces vortices, or spiral flows), play a crucial role in modulating WSS, and possibly influencing aneurysm stability \cite{isoda2010vivo, review-rayz2020}. Our results in Fig. \ref{fig:streamlines} and Fig. \ref{fig:dominant-modes} affirm that the complexity of flow near peak systole is predominantly controlled by the inflow jet, which is also affected by patient-specific morphology. Previous works classified inflow jet patterns into two types: (i) concentrated; and (ii) diffused types \cite{futami2016inflow, LE2021110238}.  Our results in Fig. \ref{fig:dominant-modes} indicate that the inflow jet of patients 3, 4, 5, and 6 is concentrated and retained more energy during wall impingement in comparison to the diffused ones (Patient 1 and Patient 2). The inflow jet's directionality and intensity are significantly dictated by the angle between the parent artery and the aneurysm's axial orientation as shown in Fig. \ref{fig:streamlines}. The comparison of flow patterns at peak systole between Patient 1 and Patient 2 (both from Group 1) also reveals that the rotational axis of the main vortex is highly sensitive to variations in the incoming flow angle and the aneurysm's aspect ratio. Our results thus further emphasize the importance of the incoming flow angle as observed in clinical practice \cite{lin2012differences}.

Broadly, there are two locations of the inflow jet's impingement: (i) distal wall (Patient 1, 2, 3, 5, and 6); and (ii) dome wall (Patient 4). The main difference in these two types resides in the pseudo-spectrum in Fig. \ref{fig:all-temporal} and Fig. \ref{fig:selected_tempmag}, which suggests that dome impingement might lead to higher flow complexity.  As illustrated in Fig. \ref{fig:peak_modes}, the impingement leads to the formation of small-scale vortex structures surrounding the inflow jet, which leads to high-frequency fluctuations. In Fig. \ref{fig:selected_tempmag}, it is evident that the prominent peaks in the high-frequency region for Patient 2 (diffused) and Patient 6 (concentrated) are $50.90$ Hz and $76.93$ Hz, respectively. 

\subsection{High-frequency Fluctuations}
\begin{figure}
	\centering
	\includegraphics[width=\textwidth]{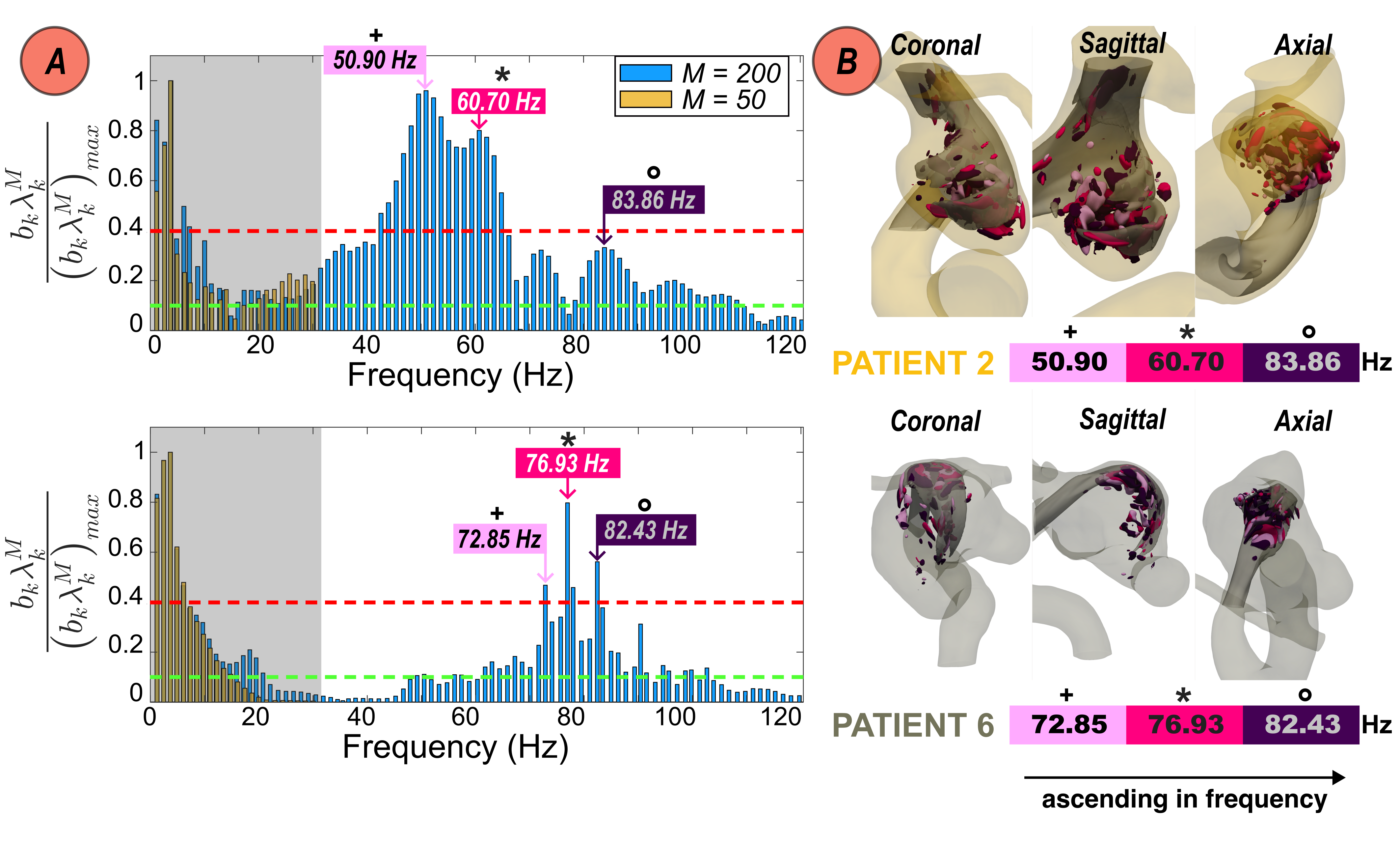}
	\caption{Group of individuals (patients 2 and 6) having significant contributions from high-frequency fluctuations in Hankel DMD analysis. (A) Pseudo-spectra shows highly stable modes in the domain $45-65$ Hz in Patient 2, and the domain $75-95$ Hz in Patient 6. (B) The 3D spatial extents of representative high-frequency modes. For Patient 2: $f_{+}=50.90$ Hz, $f_* =60.70$ Hz, and $f_o = 83.86$ Hz; For Patient 6: $f_{+}=72.85Hz$, $f_* =76.93Hz $, and $f_o = 82.43Hz$. The sensitivity analysis of Hankel DMD analysis is shown at two different temporal resolutions $\Delta \tau = 16.8$ ms ($M = 50$, yellow bars) and $\Delta \tau = 4.2$ ms ($M = 200$, blue bars). The inflow jet is visualized in (B) as a grey iso-surface using the first three dominant modes (see Figure \ref{fig:dominant-modes}. The modes are reconstructed using the ratio $\frac{\mathbf{V}}{\mathbf{V}_{max}} = 0.5$.}
	\label{fig:peak_modes}
\end{figure}

A critical factor in cavity-like flows is the formation of the primary vortex \citep{le2010pulsatile, zhang.2022} and its interaction with the distal wall resulting in high-frequency fluctuations. The presence of high-frequency fluctuations within IAs has been a subject of research due to its potential implications for aneurysm behavior and rupture. Clinically, these fluctuations manifest within a frequency band of $10 - 400$ Hz, indicative of underlying turbulence or mechanical oscillations \cite{ferguson1970turbulence, mast1995theory}. Despite advances in imaging and computational modeling, the precise origins of these high-frequency oscillations remain unclear \cite{bruneau2023understanding}. In Figures \ref{fig:streamlines} and \ref{fig:peak_modes}, our results indicate that the formation and disruption of vortex rings within the aneurysm sac might lead to high-frequency fluctuations. Our results in Fig. \ref{fig:dominant-modes} and \ref{fig:selected_tempmag} reveal that the observed high-frequency fluctuations ($60-80Hz$) are located in the vicinity of vortex ring formation and the inflow jet impingement. Our result agrees with previous \textit{in vitro} data \cite{zhang.2022} which demonstrated the link between vortex impingement and transition to turbulence. This transition induces temporal variations in WSS \cite{khan2021prevalence, le2010pulsatile, baek2009flow, ford2012exploring, valen2013high} which could influence the risk of aneurysm rupture \cite{lasheras2007biomechanics, le2013vortex}. 

Our findings, as depicted in Fig. \ref{fig:selected_tempmag}, demonstrate that high-frequency fluctuations within the range of $20-120$ Hz occur in both small and large aneurysms, regardless of aneurysm size, exemplified by Patient 2 and Patient 6, respectively. These frequencies significantly exceed those of the inflow waveform, which are limited to $1.2$ Hz, $2.4$ Hz, and $3.6$ Hz, indicating that these high frequencies result from the dynamic interactions between the inflow jet and the aneurysm sac. Specifically, these frequencies emerge predominantly around the areas where the inflow jet impinges on the aneurysm walls, as detailed in Fig. \ref{fig:peak_modes}. This spatial correlation underscores the critical impact of jet-wall interactions in generating these high-frequency fluctuations, which may influence the mechanical stress experienced by the aneurysm wall and, consequently, its structural integrity.

\subsection{Modal Analysis of Cardiovascular Flows}
Modal analysis, a branch of machine learning techniques, has gained significant attention for its application in cardiovascular flow studies due to its robust capability to dissect complex fluid dynamics \cite{taira2017modal, Arzani.2020, arzani2022machine}. Despite its potential, the incorporation of specific modal analysis techniques like DMD into cardiovascular research remains relatively under-explored \cite{csala2022comparing}. Traditionally, efforts in modal analysis have primarily focused on reconstructing flow dynamics to better understand various fluid behaviors \cite{CHATPATTANASIRI2023111759, norouzi2021flow, yu2022application}. Our previous research has demonstrated the effectiveness of DMD in parsing spatial-temporal patterns within fluid flows, thus providing deeper insights into the dynamic interactions within cardiovascular systems \cite{LE2021110238}. 

\begin{figure}
	\centering
	\includegraphics[width=\textwidth]{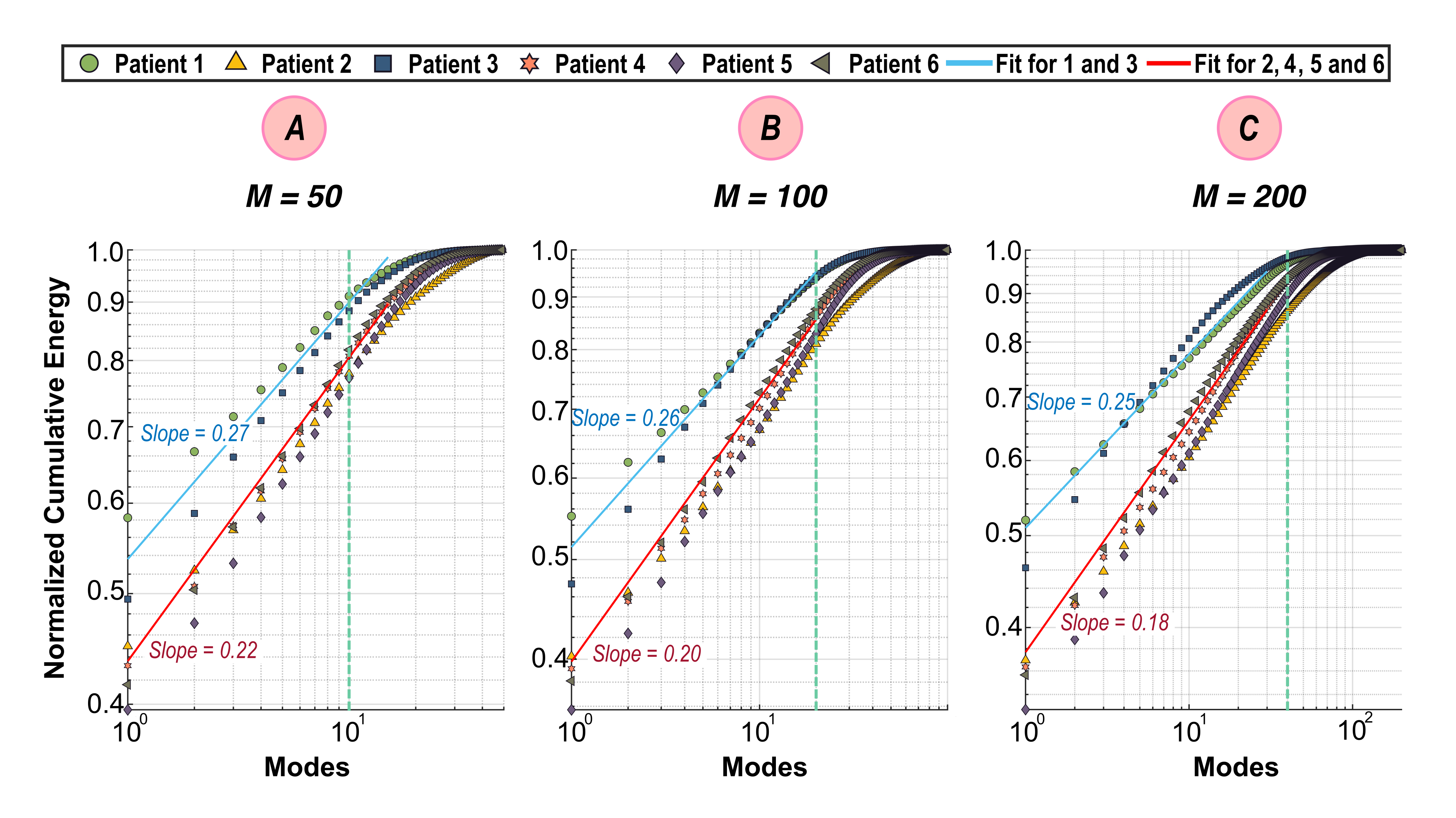}
	\caption{The stratification of flow complexities based on the cumulative energy (CE) curves from singular values of Hankelized data matrix $\mathbf{X}_{H(d=1)}$ across different temporal resolutions: (A) $M = 50$; (B) $M = 100$; and (C) $M = 200$. There is no truncation applied to the singular matrix. The CE curves are shown in a log-log scale for legibility purposes. Two separate groups can be identified consistently across all values of $M$: (1) Laminarized flow (patients 1 and 3) - blue line ($slope \approx 0.26$); and (2) Transient dynamics (patients 2, 4, 5, and 6) - ($slope \approx 0.20$). A cyan dashed line marks the accumulation at 20\% the total number of modes.}
	\label{fig:all-temporal}
\end{figure}

In our current study, the ability of CE curves to consolidate the evolution of flow fields into a single, comprehensible curve for each patient-specific scenario underscores its robustness in characterizing cardiovascular flows as depicted in Fig. \ref{fig:all-temporal}. Each mode's contribution within these curves represents the specific flow energy at that frequency, encapsulating the essence of flow dynamics within the cardiovascular system. Laminar flows are characterized by lower-frequency oscillations. The decay rate of CE curve demonstrates that it can reflect the underlying large-scale flow structures (modes with frequencies below  $10$ Hz) whereas turbulent flows exhibit a broader spectrum of high-frequency fluctuations \cite{csala2022comparing}. As shown in Fig. \ref{fig:all-temporal}, 80\% of the total flow energy across all patients can be represented with the first 20\% of total DMD modes. This finding is consistent with our previous research \cite{LE2021110238}, which has identified cumulative energy as a valuable indicator for assessing flow complexity and potentially predicting areas prone to instability within brain aneurysms. Moreover, the cumulative energy curve is highly robust and insensitive to both low and high spatio-temporal resolutions as seen in Fig. \ref{fig:spatial} and Fig. \ref{fig:all-temporal}. The cumulative energy curve varies within $1.5\%$ at all spatial resolutions $1X, 2X, 4X, 8X$ for both small (Patient 1) and large (Patient 6) aneurysms as seen in Fig. \ref{fig:spatial}. In addition, the contribution of the first mode only decreases from 0.58 at low ($M = 50$) to 0.52 at high ($M = 200$) temporal resolution for Patient 1 as seen in Fig. \ref{fig:all-temporal}. Therefore, the CE curve could be used to characterize the hemodynamics of the aneurysm sac even at low resolutions $\Delta x \approx 1.0$ mm (Table \ref{tab:spatial-resolution}) and $\Delta t \approx 20$ ms (Fig. \ref{fig:all-temporal}). 

\begin{figure}
	\centering
	\includegraphics[width=\textwidth]{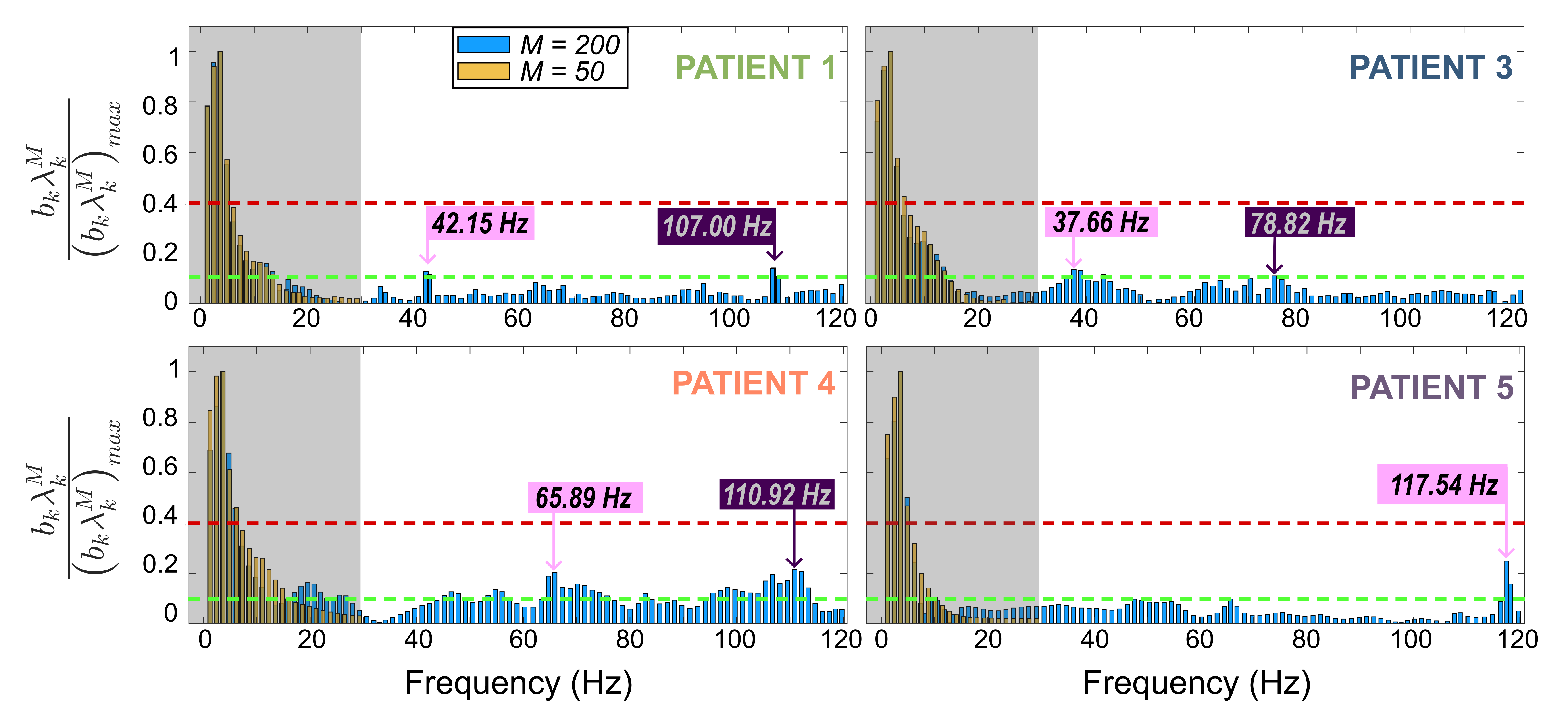}
	\caption{Pseudo-spectra from a group of individuals (patients 1, 3, 4, and 5) with low contributions from high-frequency fluctuations in Hankel DMD analysis. The sensitivity analysis of Hankel DMD analysis is shown at two different temporal resolutions $\Delta \tau = 16.8$ ms ($M = 50$, yellow bars) and $\Delta \tau = 4.2$ ms ($M = 200$, blue bars). The temporal magnitudes of modes are consistent values between $M = 50$ and $M = 200$ at lower resolution (grey zone). Note that the first mode (time-averaged) is excluded from the analysis. Data for high-frequency fluctuations are only available for $M = 200$. The dashed lines indicate the constant normalized magnitude of $10\%$ (green) and $40\%$ (red), respectively.}
	\label{fig:selected_tempmag}
\end{figure}

Based on the slope of the CE curves as shown in Fig. \ref{fig:all-temporal}, there are two patient groups: (i) "Laminarized flows" (Patient 1 and Patient 3, slope $ \approx 0.26)$; and (ii)  "Transient dynamics" (Patient 2, Patient 4, Patient 5, and Patient 6, $slope \approx 0.2$). Upon revisiting the frequency spectra in Fig. \ref{fig:selected_tempmag}, it becomes evident that high-frequency modes are the main factors driving the slopes down at different rates. From $M = 50$ to $M = 200$, the number of modes quadruples, from 50 to 200 modes. The corresponding frequency ranges increase from $30$ Hz to $120$ Hz. In "Laminarized flows" (Patient 1 and Patient 3), the averaged magnitude of high-frequency modes is less than 10\% of the most dominant mode ($3.63$ Hz). On the contrary, high-frequency modes within the "Transient dynamics" group get higher magnitudes and contribute significantly to the total dynamics.  

As illustrated in Fig. \ref{fig:selected_tempmag}, DMD has proven invaluable in identifying and differentiating high-frequency dynamics across different aneurysm sizes. This analytical strength of DMD extends beyond the frequency analysis to offer a detailed mapping of the spatial distribution of these modes. In particular, this contribution is most pronounced in Patient-2 and Patient 6 as seen in Fig. \ref{fig:selected_tempmag}.  In Patient 2, the magnitude of higher frequencies (from $42$ Hz to $68$ Hz) is significantly large. Specifically, the peak frequency at $50.9$ Hz contributes as large as the most dominant mode at a much lower frequency ($3.81$ Hz). DMD also demonstrates its ability to locate spatially both the large-scale (low-frequency) and small-scale (high-frequency) flow structures as displayed in Figures \ref{fig:dominant-modes} and \ref{fig:peak_modes}. This capability is instrumental for pinpointing potential regions within the cardiovascular system that are prone to instabilities and may transition into turbulent states. Such regions are often precipitated by complex inflow jet dynamics, suggesting that DMD can be used as a tool in predicting areas of potential flow disruption and aiding in the strategic planning of therapeutic interventions \cite{LE2021110238}.

DMD is renowned for its ability to parse high-dimensional data into coherent structures or modes, facilitating the understanding of complex fluid dynamics \citep{taira2017modal}. However, DMD is susceptible to noise and occasionally generates spurious modes \cite{schmid2022,cohen2021examining}. To address these challenges, the Optimized DMD method has been developed, which refines the mode identification process by incorporating an optimization framework directly into the mode computation. This innovation significantly mitigates the discrepancies typically observed with traditional DMD applications \cite{askham2018variable}. 

In our analysis, a rigorous evaluation of the performance of both Hankel DMD and Optimized DMD techniques is carried out for flow dynamics within brain aneurysms. Our results in Fig. \ref{fig:patient4_optDMD} show that Optimized DMD results in a greater proportion of decayed modes (modes are within the unit circle). Thus, the Optimized DMD method suppresses unstable modes and concentrates flow energy into more stable, lower-frequency modes. This characteristic is prominently displayed in Fig. \ref{fig:patient4_optDMD} for Patient 4 when the modes are compared with ones obtained by the Hankel DMD technique (see also Figures \ref{fig:dominant-modes}, \ref{fig:selected_tempmag}, \ref{fig:peak_modes}). Optimized DMD modes predominantly occupy the lower frequency band (below $15$ Hz), and higher frequency fluctuations are effectively absent. While this feature of Optimized DMD enhances its utility in pinpointing areas of flow instability such as regions of jet impingement, it also restricts the capability to discern the forcing frequencies from the transient modes. Given these observations, Optimized DMD might be suitable for identifying turbulent regions in cardiovascular flows. However, care must be taken in interpreting the obtained Optimized DMD frequencies.

\subsection{Limitation}
One of the primary challenges in patient-specific modeling is the accurate replication of \textit{in vivo} conditions due to a lack of comprehensive boundary condition data such as tissue properties, inlet flow waveforms, and outlet resistances \cite{taylor2009patient}. First, a generic flow waveform and a fully developed flow condition are set for the inlet and outlet conditions, respectively, in our study. This approach may not accurately reflect the patient-specific hemodynamic environment \cite{review-rayz2020}. Second, a technique was employed to isolate the ICA from the CoW. The terminal segment was modified which can introduce uncertainties by creating anatomical variation from the authentic version in the DICOM image. When dealing with multiple outlets, more accurate approaches include utilizing the Murray's law, or incorporating the aorta-to-cerebral vasculature model with three-element Windkessel model for multiple outlets \cite{fetalPCA}. However, the evolution of Murray's law has not attained full maturity \cite{generalizingMurray}. In addition, the outlet resistances cannot be measured from experiments and they are often tuned for patient-specific cases. Finally, the use of rigid wall boundary conditions could influence the dynamics of flow, particularly near the wall surfaces where flow instabilities occur. These limitations necessitate further investigations into the links between aneurysm locations (ICA, ACA and MCA) or wall compliance \cite{bruneau2023understanding} and high-frequency fluctuations of hemodynamics.

While DMD offers powerful insights into fluid dynamics by parsing high-dimensional data into coherent structures, it is inherently limited by its linear nature and the potential generation of spurious modes. DMD is associated with several well-documented limitations, such as its inherent linear approximation nature, time-forward bias, and the generation of "false modes" when dealing with noisy data. These "false" modes can arise due to noise in the data or from complex, transient dynamics that are not fully captured by linear models \cite{LE2021110238}. The linear approximation inherent in DMD and the challenges in distinguishing transient from persistent phenomena necessitate careful interpretation of the results, especially when considering clinical applications. As seen in Fig. \ref{fig:domain-impact}, the choice of VOI for input data influences the cumulative energy curves the most. Thus, it is essential to choose VOI appropriately to minimize input noise in practice. To enhance the reliability of DMD methodologies in clinical settings, future research should explore larger datasets and more diverse patient cohorts to validate the robustness of these computational techniques across a broader spectrum of aneurysm cases. This expansion would help mitigate the risk of overfitting to specific data characteristics and improve the model's predictive power and clinical relevance.

%%=================================%
%--------- CONCLUSIONS ------------%
%==================================%
\section{Conclusions}
\label{sec:conclusions}
The current study demonstrates the potential use of Dynamic Mode Decomposition, an unsupervised machine learning, to stratify saccular aneurysms hemodynamics. First, Computational Fluid Dynamics (CFD) simulations are carried out to produce reliable data for analysis of the fluid flow in patient-specific models of aneurysms. Our CFD results provide noise-free and high-resolution flow fields ($\approx 100 ~\mu \text{m}$ and $\approx 25 ~\text{ms}$), which are much finer than available clinical data (CT or MRI). A systematic procedure of the spatiotemporal coarsening is carried out successively to generate a total of 72 cases with various spatial and temporal resolutions. Second, the Hankel DMD method can identify the dominant modes and flow instabilities inside the aneurysm sac. Our results indicate that the dominant modes bear a strong resemblance to the frequencies found in the waveform prescribed at the inlet. However, the interactions of the inflow jet with the endothelial wall and vortex structures lead to the formation of high-frequency modes. Finally, based on the pseudo-spectra, aneurysmal hemodynamics can be categorized into two groups: (i) "Laminarized flows"; and (ii) "Transient dynamics" based on the slope of the cumulative energy curve. In summary, our main conclusions are:
\begin{enumerate}    
	\item \textit{Patient-specific flow dynamics in ICA aneurysms:} DMD analysis reveals unique 3D flow structures in each patient, independent of aneurysm size or aspect ratio.
	\item \textit{Robust flow characterization at low resolutions:} Energy pseudo-spectrum effectively captures aneurysmal flow dynamics across various spatial and temporal resolutions.
	\item \textit{Potential tool for clinical rupture risk assessment:} DMD offers a new quantitative approach (pseudo-spectra, cumulative CE curve) for evaluating aneurysm severity.
\end{enumerate}

This study endeavors to delve deeper into the challenges and opportunities presented by the traditional DMD in the context of IAs and to validate the efficacy of two DMD techniques (Hankel and Optimized) for aneurysmal flows. Our results indicate that the Hankel DMD method is an appropriate tool for determining flow complexities even at relatively low temporal and spatial resolutions. 

%%=================================%
%------- ACKNOWLEDGEMENT ----------%
%==================================%
\begin{acknowledgments}
We acknowledge the financial support from the New Discoveries in the Advanced Interface of Computation, Engineering, and Science (ND-ACES) project (National Science Foundation, NSF \#1946202). The computational work was performed using the computational resources of the Center for Computationally Assisted Science and Technology (CCAST) at North Dakota State University enabled by NSF MRI \#2019077 and the Extreme Science and Engineering Discovery Environment (XSEDE) Bridges at the Pittsburgh Supercomputing Center under the allocation CTS200012. The second author, Davina Kasperski, was funded by a fellowship from the Biomedical Engineering Program, College of Engineering, North Dakota State University.
\end{acknowledgments}

%%=================================%
%----------- CLOSURES -------------%
%==================================%
\section*{Data Availability Statement}
The data that support the findings of this study are available from the corresponding author upon reasonable request.

\section*{Author Declarations}
\subsection*{Conflict of Interest}
The authors have no conflicts to disclose.

\subsection*{Author Contribution}
\textbf{Thien-Tam Nguyen}: Data curation (lead); Formal analysis (lead); Investigation (lead); Methodology (lead); Software (lead); Visualization (lead); Writing – original draft (lead); Writing – review \& editing (equal). \textbf{Davina Kasperski}: Data curation (supporting); Formal analysis (supporting); Investigation (supporting); Methodology (supporting); Writing – original draft (supporting). \textbf{Trung Bao Le}: Conceptualization (lead); Methodology (equal); Supervision (lead); Writing – original draft (supporting); Writing – review \& editing (equal); Funding acquisition (lead). \textbf{Phat Kim Huynh}: Conceptualization (supporting); Methodology (equal); Writing – review \& editing (supporting). \textbf{Trung Quoc Le}: Conceptualization (supporting); Methodology (supporting).

%%=================================%
%---------- REFERENCES ------------%
%==================================%
\bibliography{aip_main}

%==================================%
%----------- TABLES ---------------%
%==================================%
\clearpage

\begin{table}[h]
	\centering
	\begin{tabular}{|c|c|c|c|c|c|c|c|c|c|c|}
		\hline
		\hline
		Patient No. & AneuRisk ID & $D$ & $W$ & $H$ & AR & $Re_{max}$ & $\alpha$ & W/D & An & Description \\
		\hline
		1 & C0002 & 3.6 &  4.3 &  7.7 & 1.79  & 537 & 2.7 & 1.2 & 2.7 & Small						\\
		2 & C0014 & 4.5 &  6.9 &  5.2 & 0.75  & 671 & 3.3 & 1.5 & 3.5 & Small						\\
		\hline
		3 & C0042 & 3.9 &  8.4 &  7.7 & 0.92  & 582 & 2.9 & 2.1 & 4.9 & Medium						\\
		4 & C0016 & 3.0 &  7.7 &  7.4 & 0.96  & 447 & 2.2 & 2.5 & 5.9 & Medium						\\
		\hline
		5 & C0036 & 3.9 & 10.7 & 14.2 & 1.32  & 582 & 2.9 & 2.7 & 6.3 & Large + bleb				\\
		6 & C0075 & 3.5 & 12.0 & 11.8 & 0.98  & 522 & 2.6 & 3.4 & 7.9 & Large + bleb				\\
		\hline
		\hline
	\end{tabular}
	\caption{Patient ID and anatomical attributes of corresponding aneurysms from the \href{https://github.com/permfl/AneuriskData}{AneuRisk project}. All aneurysms are located at the Internal Carotid Arteries. The dimensions of the aneurysms ($D$, $W$, and $H$) are defined in Fig. \ref{fig:morphology}. The peak Reynolds number at the inlet is based on the bulk flow velocity $U_0 = 0.5m/s$ as $Re_{max} =\frac{U_0 D}{\nu}$. The Wormersley number is defined as: $\alpha = \frac{D}{2} \sqrt{\frac{2 \pi}{T \nu}}$. Blood viscosity is $\nu = 3.35 \times 10^{-6} m^2/s$. The Aneurysm Number is defined as $An = PI \frac{W}{D}$. The pulsatility index ($PI$) is computed from the pulsatile flow as shown in Fig. \ref{fig:setup}. Acronyms: Artery Diameter ($D$), Aneurysm Diameter ($W$), Aneurysm Height ($H$). Aspect Ratio (AR) is defined as the ratio between the aneurysm's height over the neck width.The pulsatility index $PI = \frac{U_{max} - U_{min}}{\overline{U}} = 2.3$ (from the inlet waveform).}
	\label{tab:anatomical-parameters}
\end{table}

\clearpage

\begin{table}[b]
	\centering
	\begin{tabular}{|c|c|c|c|c|c|}
		\hline
		\hline
		Spatial  &   Patient & \multicolumn{3}{|c|}{ Resolution (mm)} & Average \\
		\cline{3-5}
		Coarsening Factor &  No.  & $\Delta x$ &  $\Delta y$ &  $\Delta z$ & (\textit{mm}) \\
		\hline
				&   1  	&   0.09   &   0.12   &   0.12   &            			\\ 
				&   2   &   0.11   &   0.16   &   0.16   &             			\\ 
		$1X$    &   3   &   0.11   &   0.10   &   0.13   &   0.12      			\\ 
				&   4   &   0.06   &   0.09   &   0.10   &             			\\ 
				&   5   &   0.14   &   0.14   &   0.20   &             			\\ 
				&   6   &   0.10   &   0.13   &   0.14   &             			\\
		\hline
				&   1   &   0.17   &   0.25   &   0.23   &             			\\
				&   2   &   0.22   &   0.31   &   0.32   &             			\\ 
		$2X$    &   3   &   0.22   &   0.20   &   0.25   &   0.24     			\\ 
				&   4   &   0.13   &   0.17   &   0.19   &             			\\ 
				&   5   &   0.29   &   0.27   &   0.41   &             			\\ 
				&   6   &   0.19   &   0.27   &   0.28   &             			\\
		\hline
				&   1   &   0.35   &   0.50   &   0.46   &             			\\
				&   2   &   0.45   &   0.63   &   0.63   &             			\\ 
		$4X$    &   3   &   0.45   &   0.40   &   0.50   &   0.49      			\\ 
				&   4   &   0.26   &   0.35   &   0.38   &             			\\ 
				&   5   &   0.58   &   0.55   &   0.82   &             			\\ 
				&   6   &   0.39   &   0.54   &   0.56   &             			\\ 
		\hline
				&   1   &   0.70   &   0.10   &   0.93   &             			\\
				&   2   &   0.90   &   1.26   &   1.27   &             			\\ 
		$8X$    &   3   &   0.90   &   0.80   &   1.00   &   0.98      			\\ 
				&   4   &   0.52   &   0.70   &   0.76   &             			\\ 
				&   5   &   1.15   &   1.09   &   1.64   &             			\\ 
				&   6   &   0.77   &   1.07   &   1.11   &             			\\ 
		\hline
		\hline
	\end{tabular}
	\caption{Spatial resolution inside the aneurysm sac for all cases resulted from the spatial resampling technique. The spatial resolution of $1X$ is the Computational Fluid Dynamics (CFD) result (no coarsening). The resampling rates of $2X$, $4X$, and $8X$ are the factors applied when performing coarsening on the results from the original CFD data. The sampling rate $8X$ has a spatial resolution in a similar order to the ones from MRI data.}
	\label{tab:spatial-resolution}
\end{table}

%%=================================%
%----------- FIGURES --------------%
%==================================%
\clearpage

\clearpage

\clearpage

\clearpage

\clearpage

\clearpage

\clearpage

\clearpage

\clearpage

\clearpage

\clearpage

\clearpage

\clearpage

\end{document}